%% file: aaai-anonymous-submission-latex-2024.tex
\documentclass[letterpaper]{article} 

\usepackage{aaai24}  

\usepackage{times}  
\usepackage{helvet}  
\usepackage{courier}  
\usepackage[hyphens]{url}  
\usepackage{graphicx} 
\usepackage{natbib}  
\usepackage{caption} 
\usepackage{algorithm}
\usepackage{algorithmic}

\usepackage{newfloat}
\usepackage{listings}
\DeclareCaptionStyle{ruled}{labelfont=normalfont,labelsep=colon,strut=off} 
\floatstyle{ruled}
\newfloat{listing}{tb}{lst}{}
\floatname{listing}{Listing}
\usepackage{fancyvrb}
\usepackage{multirow}
\usepackage{booktabs}       
\usepackage{subcaption}
\usepackage{enumitem}
\usepackage{natbib}
\usepackage{color, colortbl}
\definecolor{LightCyan}{rgb}{0.88,1,1}
\definecolor{Gray}{gray}{0.9}
\input{math_commands.tex}

\newcommand{\probability}[1]{\mathbb{P}\left[ #1 \right]}
\definecolor{codegreen}{rgb}{0,0.6,0}
\definecolor{codegray}{rgb}{0.5,0.5,0.5}
\definecolor{codepurple}{rgb}{0.58,0,0.82}
\definecolor{backcolour}{rgb}{0.95,0.95,0.92}
\newcommand\doesnotentail{\mkern-2mu\not\mkern2mu\vdash}
\newcommand\system{\texttt{speculyzer}}
\usepackage{adjustbox}
\usepackage{fvextra}
\title{Toward Trustworthy Neural Program Synthesis}
\author {
    Wen-Ding Li\equalcontrib\textsuperscript{\rm 1},
    Darren Key\equalcontrib\textsuperscript{\rm 1},
    Kevin Ellis\textsuperscript{\rm 1}
}
\affiliations {
    \textsuperscript{\rm 1}Department of Computer Science, Cornell University, USA\\
    wl678@cornell.edu, dyk34@cornell.edu, kellis@cornell.edu
}

\usepackage{bibentry}

\begin{document}

\maketitle

\begin{abstract}
We develop an approach to estimate the probability that a program sampled from a  large language model is correct. Given a natural language description of a programming problem, our method samples both candidate programs as well as candidate predicates specifying how the program should behave. This allows learning a model that forms a well-calibrated probabilistic prediction of program correctness.
Our system also infers the which predicates are useful to explain the behavior of the generated code, and humans preferred these in a human study over raw language model outputs. Our method is simple, easy to implement, and maintains state of the art generation accuracy results.
\end{abstract}

\section{Introduction}
Picture a future where AI systems attempt to close GitHub issues by generating source code given only the natural language of the GitHub issue.
Such systems might not come out next year, but when or if they ever do, they will likely leverage large neural network language models for source code~\citep{codex, austin2021program}.
These neural systems are good, but not perfect.
Suppose 75\% of the time, such systems propose a correct fix to the GitHub issue.
The other 25\% of the time, they produce plausible looking code containing subtle bugs.
Would you use this system?

Most engineers would be reluctant to use such a system, because it fails to build trust with the user.
When it fails, it cannot detect its own failure.
When it succeeds, it doesn't present a human-understandable explanation of why its program behaves as intended.
In this paper we seek steps towards rectifying this lack of trust by building natural-language conditioned program synthesizers that are more trustworthy in several complimentary ways:
\begin{itemize}[leftmargin=*]
    \item \textbf{Calibration}: We want systems that, when they cannot solve a programming problem,  simply return no answer, rather than return a (possibly subtly) incorrect program. We conjecture that it is better to fall back on the human programmer, rather than risk introducing bugs.
    Contrast the situation with natural language translation: Unlike natural language, programs are brittle, and more time-consuming to look into and understand. And debugging bad code,  unlike proofreading language, can be harder than just writing it yourself. We will accomplish this by having a classifier that predicts whether the program is correct, and ensuring that the classifier is well calibrated~\cite{platt1999probabilistic, kuhn2023semantic}.
    \item \textbf{Explainability}: To help humans understand the output of a neural network that is writing code, we want a system that can explain its outputs by generating informative
    and interpretable checks on program behavior.
    We propose a characterization of what makes an explanation of program behavior `good', and validate in a human user study that this characterization produces better explanations than a raw large language model by itself.
    \item \textbf{Accuracy}: Ideally, trustworthy systems should be more accurate, solving more programming problems.
    This goal might seem to be in tension with the previous two.
    Surprisingly, we find our methods also boost overall accuracy on natural language to code generation problems.
\end{itemize}

\begin{figure*}[t]
\center
\includegraphics[width = 0.8\textwidth]{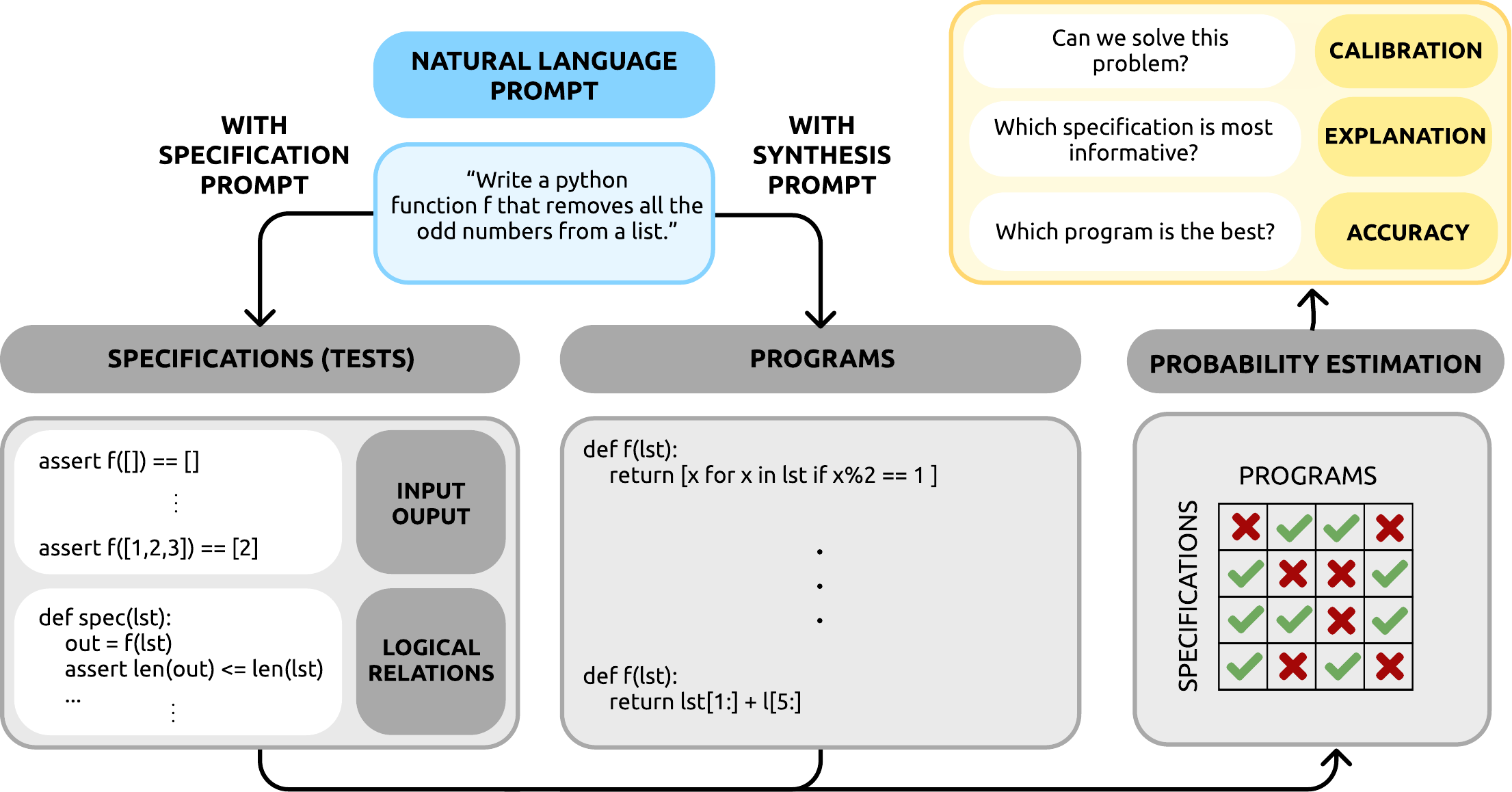}
\caption{Our \system{} system inputs a natural language description of a programming problem.
It uses large language models to independently sample candidate programs, and candidate specifications of what the program should do.
Because natural language is informal, we cannot mechanically check programs against it, but logical relations and input-outputs can be mechanically checked against.
The result of this validation step is fed to a learned model which predicts whether the problem can be solved; if so, which program is correct; and which specs best explain the behavior of the program, am would be useful for judging whether that program is correct or incorrect.
}\label{fig:overview}
\end{figure*}

Our high-level approach has a neural network propose candidate program solutions and independently propose predicates that correct solutions should satisfy, such as Input/Output tests, known as \textbf{specifications} (`specs', Fig.~\ref{fig:overview}).
Although specs can refer to a broad range of ways to specify the behavior of programs, here we only consider two kinds: (1) input-output test cases, and (2) parameterized logical relation tests which execute the program and check some relation between input and output.
In general, a spec can be any mechanically checkable property.
We check the programs against the specs, and learn to use this checking to predict if the system knows how to solve the problem at all, and if so, which program(s) are probably the right solution.
Intuitively, we ask the language model to `check its work' by generating specs.
We call our approach \system{}, short for `Specification Synthesizer', because in addition to synthesizing programs, it synthesizes specs.

Our work makes the following contributions:
\begin{enumerate}[leftmargin=*]
    \item \textbf{Calibration} A method to give a well-calibrated probabilistic estimate of whether a program is correct which enables analysis of metrics connected to trust and safety
    \item \textbf{Explainability} A method for identifying the specifications most likely be useful to humans as an explanation of program behavior, and a validation of that approach in a human study
    \item \textbf{Accuracy} Demonstration that the above contributions do not impair the overall accuracy of programs synthesizers, and can sometimes let them solve more problems overall
\end{enumerate}


\section{Related Work}

\textbf{Program synthesis.} Automatically constructing software has been a longstanding goal of computer science~\citep{1702636,gulwani2017program}.
Classic program synthesizers input a formal specification, and then either search or logically derive a program guaranteed to satisfy that formal specification~\citep{alur2013syntax}.


\noindent\textbf{Large language models for source code.}
Our work uses large language models for source code~\citep{codex, austin2021program}.
These neural networks generate source code conditioned or `prompted' by a mix of code and natural language (the natural language is usually represented as code comments).
Such models are typically implemented as large transformers~\citep{transformers,gpt}. 

Following the introduction of large transformer-based language models for source code, there has been work on how to boost the accuracy of those models.
Here, accuracy means the probability of sampling a correct program conditioned on a natural-language prompt.
Accuracy is often measured by functional correctness with the \emph{pass@k} metric, which considers drawing $k$ IID samples from the language model and testing if any of those samples pass a set of holdout test cases.
Toward boosting \emph{pass@k}, researchers have considered \emph{clustering} sampled programs according to the outputs they produce on test inputs~\citep{shi2022natural, li2022competition}.
For example, AlphaCode prioritizes large `clusters' of samples with the exact same input-output behavior~\citep{li2022competition}, effectively reranking the samples from the language model according to how likely they are to solve the task.
A complementary reranking strategy is to train a second neural network to predict program correctness, as explored in~\cite{inala2022fault}.

The closest work to ours is CodeT~\citep{chen2022codet}, which also  generates programs as well as input-output test cases, with the goal of boosting \emph{pass@k}.
The qualitative difference between our systems is that we designed \system{} to build trust in a variety of ways by synthesizing specifications, centered around first forming well-calibrated probability estimates, only boosting \emph{pass@k} as a side effect.
We also incorporate input-output test cases as a special case of specs in general.


Engineering safe, trustworthy language models has received considerable attention by the AI safety~\citep{lambda} and AI alignment communities~\citep{kadavath2022language}.
These works find that one can train classifiers which predict the truthfulness or safety of language outputs by inspecting the hidden activations of the model or even by simply `asking' the model if its output is correct or safe.
We see this family of efforts as complementary: For programs, it is possible to formally specify correctness properties, which is not generally true in NLP, so we focus on formal properties (specifications) here.
Nonetheless, one can train statistical predictors of program correctness~\citep{inala2022fault}.
Broadly however, we think that program synthesis offers unique opportunities for building trust through symbolic methods.
Although statistically reranking language model outputs via a second neural network improves raw performance, we believe it is a suboptimal trust-builder: an inscrutable neural network cannot guarantee the correctness of another inscrutable network. Here we advocate that properties which are symbolically checkable and human-comprehensible should play a role, and examine certain specifications as basic examples of such properties.



\section{Methods}

Given a natural-language prompt describing a programming problem, our goal is to produce a well-calibrated estimate probability of each program sample from language model being correct.
Our approach independently samples a set of candidate programs $\mathcal{P}$ and a set of candidate specs $\mathcal{S}$. 
Specs can be either input-output testcases, or logical relations (Fig.~\ref{fig:overview}).
We write $\mathcal{T}$ for the set of test cases and $\mathcal{R}$ for the set of logical relations, so $\mathcal{S}=\mathcal{T}\cup \mathcal{R}$.
Each program $p\in \mathcal{P}$ is checked against each spec $s\in \mathcal{S}$, and basic statistics of program-spec agreement are computed.
These statistics are aggregated by a learned model into an  estimated probability that the program is correct.
Programs whose probability falls below a threshold are discarded.
Any remaining programs are sorted by probability and returned to the user as possible solutions, together with certain specs they pass.
Returning specifications allows the user to check that the code has the intended behavior.
This architecture lets the system learn how to predict when it cannot solve a problem, and also learn to rank candidate solutions and their corresponding specs.



\subsection{Sampling programs and specs}
Given a string \texttt{prompt} describing a programming problem, we sample $n=100$ candidate programs (the set $\mathcal{P}$) and candidate specs (the set $\mathcal{S}$).
Both sets are sampled using a pretrained language model, which can probabilistically generate program source code conditioned on a prompt.
We write $P^\text{LM}(\cdot |\texttt{prompt})$ for the conditional distribution over programs, given \texttt{prompt}.
If a program $p\in \mathcal{P}$, then $p\sim P^\text{LM}(\cdot |\texttt{prompt})$.
To sample specs, we deterministically transform the prompt as in Fig.~\ref{fig:prompting} and Appendix
, then draw iid samples from the language model to construct relations $\mathcal{R}$ and input-output test cases $\mathcal{T}$.

One motivation for generating logical relations is that large language models are famously bad at predicting the output of a novel program~\cite{DBLP:journals/corr/abs-2112-00114}.
However, given a generic input, they can produce a variety of reasonable constraints on that the output should obey, if they are prompted in a program-of-thought style~\cite{chen2022program,gao2023pal}.
Logical relations also resemble unit test harnesses, like those used in property based testing,~\cite{fink1997property} which are likely part of the model's training data.
We therefore suspected that although input-outputs are the superior form of test for typical programming tasks, logical relations could serve the long tail of novel tasks that the language model cannot reliably predict outputs for.


\begin{figure*}[h]
    \centering
    \includegraphics[width=1\textwidth]{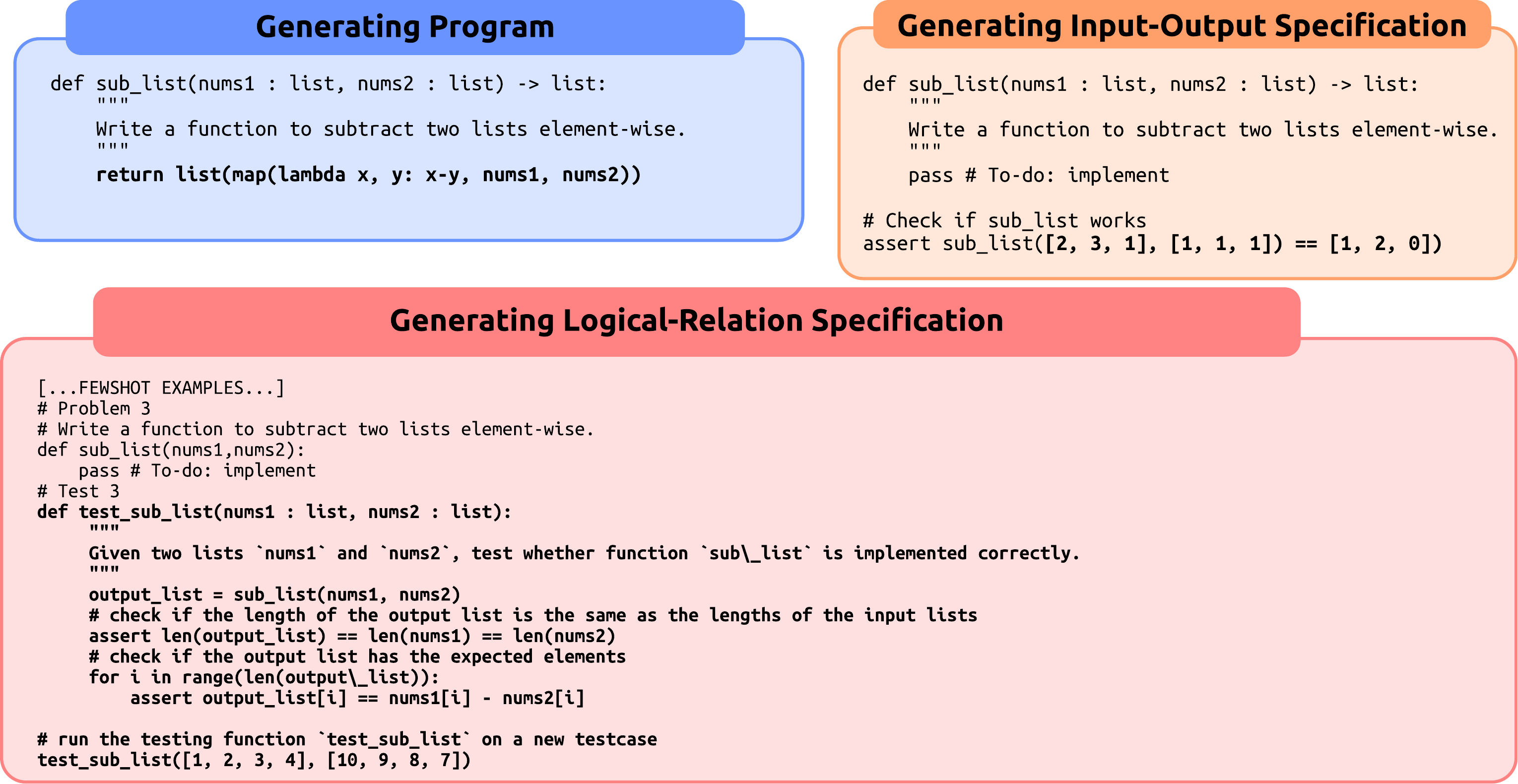}
\caption{Our systems uses different prompts to generate programs, input-output tests, and logical relations. Here we also show the example completion from the model in bold.}\label{fig:prompting}
    \label{fig:prompting}
\end{figure*}




\subsection{Checking specs against programs}
To judge the correctness of a program $p$, we would like to know whether each specification $s\in \mathcal{S}$ is actually true of $p$, notated $p\models s$.
If the specification $s$ is an input-output test, this checking is trivial: the program $p$ can be run on the input and checked to see if it yields the desired output.
But if $s$ is a logical relation, checking if $p\models s$ over the entire input space becomes undecidable.
Thus we require an approximate validation approach for logical relations.
As a heuristic approximation we ask the language model to generate a few candidate inputs on which to run the relational test, effectively using the language model as a fuzzer.
This causes us to overapproximate  the set of relations a program satisfies.
Notionally we write 
$p\vdash_T s$ to mean that spec $s$ 
is inferred to be true of program $p$ according to testing methodology $T$.
If $T=IO$, then
$p\models s$ iff $p\vdash_{IO} s$.
If $s$ is a relation, we instead have 
$p\models s$ implies $p\vdash_{Rel} s$.
To avoid confusion, we always show the concrete inputs on which a logical relation was tested.

\subsection{Scoring and analyzing test coverage}
Given programs $\mathcal{P}$ and specs $\mathcal{S}$, we produce an estimated probability for each $p\in \mathcal{P}$ that $p$ is correct.
Assuming, on average, specs correctly formalize the informal natural-language intention, satisfying more specs should increase our confidence in a program.
Additionally, if many sampled programs exhibit identical behavior on the specs, then we should increase our confidence in those programs, because this indicates high marginal probability of that behavior under $P^\text{LM}(\cdot |\texttt{prompt})$.
This `clustering' of candidate solutions according to their execution behavior, and prioritizing large clusters, has been successfully used by AlphaCode~\citep{li2022competition}, 
Minerva~\citep{lewkowycz2022solving}, and others~\citep{shi2022natural}.
It is also related to \emph{observational equivalence} from classic program synthesis~\citep{10.1145/2499370.2462174}, which treats programs as identical if they have the same outputs on test inputs.


Therefore, we estimate the probability of correctness for each program $p\in \mathcal{P}$ using a logistic regressor over features $\phi(p, \mathcal{P}, \mathcal{S})$ and learned parameters $\theta$. 
The features 
include \textbf{i/o pass rate} (input-output specs passed), \textbf{relation pass rate} (logical-relation specs passed, under fuzzing), \textbf{cluster size} (\# of other programs satisfying the same specs), the ordinal rank of the preceding features and the normalized log probability of the sampled programs:
\begin{align}
&\begin{aligned}
&\begin{aligned}
&\text{score}_\theta(p | \mathcal{P}, \mathcal{S}) =\text{Sigmoid}\left( \theta\cdot    \phi(p, \mathcal{P}, \mathcal{S})+\theta_0 \right) \\
\end{aligned}
&&
\end{aligned}\\
&\quad{\text{where the components of }\phi(p, \mathcal{P}, \mathcal{S})\text{ include the follow}} \nonumber\\
&\quad{\text{-ing and their ordinal ranks:}} \nonumber\\
&\begin{aligned}
&\quad\text{IOPass}(p, \mathcal{P}, \mathcal{T}\cup\mathcal{R})=\frac{1}{|\mathcal{T}|}\sum_{s\in \mathcal{T}}\indicator{p\vdash_\textit{IO} s}
\\
&\quad\text{RelPass}(p, \mathcal{P}, \mathcal{T}\cup\mathcal{R})=\frac{1}{|\mathcal{R}|}\sum_{s\in \mathcal{R}}\indicator{p\vdash_\textit{Rel} s}&&\nonumber\\
&\quad \text{ClusterSize}(T, p, \mathcal{P}, \mathcal{S}) \nonumber\\
&=\sum_{p'\in \mathcal{P}}\prod_{s\in \mathcal{S}}\indicator{(p\vdash_T s)=(p'\vdash_T s)} 
\text{where }T\in\left\{ IO, Rel \right\}\nonumber\\
&\quad \text{NormLogProb}(p)=\frac{1}{\text{length}(p)}\log P^{\text{LM}}(p |\texttt{prompt})
\label{eq:logisticdefinition}
\end{aligned}
\end{align}

We fit $\theta$ via maximum likelihood
on a corpus $\mathcal{D}$ containing triples $\langle\mathcal{P}, \mathcal{S}, \mathcal{G}\rangle$ of programs $\mathcal{P}$ and specifications $\mathcal{S}$, both sampled from the same program problem, and ground-truth testcases $\mathcal{G}$, which serve as a proxy for program correctness.
The ground-truth testcases $\mathcal{G}$ are assumed to be unavailable at test time, because our goal is synthesis from informal specifications like natural language.
We use gradient ascent to maximize the log likelihood, $\mathcal{L}$:
\begin{align}
\mathcal{L}=\sum_{\substack{\langle\mathcal{P}, \mathcal{S}, \mathcal{G}\rangle\in \mathcal{D}\\p\in \mathcal{P}}}\Biggl( \indicator{p\vdash_\textit{IO} \mathcal{G}}\log\text{score}_\theta(p | \mathcal{P}, \mathcal{S}) \nonumber \\
+\indicator{p\doesnotentail_\textit{IO}  \mathcal{G}}\log\left( 1- \text{score}_\theta(p | \mathcal{P}, \mathcal{S})  \right)\Biggr)
\end{align}

\subsection{Test time metrics}
\textbf{Precision-Recall.}
Ultimately our goal is a trustworthy system that proposes program solutions whenever it can, but avoids proposing buggy code.
Toward those ends, we seek high \emph{precision} without sacrificing \emph{recall}.
High precision means that when the system suggests a program, it is probably correct.
Precise systems foster trust because they don't propose wrong answers, though they may decline to provide an answer in the first place.
High recall means a correct program achieves the top rank: In other words, the system can solve a lot of programming problems, though it might make more mistakes in the process.

The tradeoff between precision and recall can be tuned by a thresholding parameter, $\tau$.
A candidate program is discarded if its score falls below the threshold $\tau$.
If all programs are discarded, the system declines to provide an output for the programming problem, and otherwise the system outputs a ranked list of programs sorted by score.

We define Precision and Recall as follows, which respectively measure (1) whether a correct program is top-ranked whenever any program scores above $\tau$ and (2) how often a correct program scoring above $\tau$ is the top ranked. Note that these definitions are generalizations of precision/recall for binary classification.
\begin{align}
&\text{Precision}@k=\frac{\text{TruePositives}@k}{\text{PredictedPositives}}\label{eq:actualpositives} \nonumber\\
&\text{Recall}@k=\frac{\text{TruePositives}@k}{\text{ActualPositives}}\nonumber\\
&\text{TruePositives}@k \nonumber\\
&=\nonumber
\sum_{\langle\mathcal{P}, \mathcal{S}, \mathcal{G}\rangle\in \mathcal{D}}
\indicator{\begin{array}{l}
     \exists p\in \mathcal{P}:p\vdash_\textit{IO}\mathcal{G} \wedge \\
     \tau\leq \text{score}_\theta(p|\mathcal{P}, \mathcal{S}) \wedge\\
      p\in \text{top-k}_{p'\in \mathcal{P}} \text{score}(p' |\mathcal{P}, \mathcal{S})\end{array}}\nonumber\\
&\text{PredictedPositives}\nonumber\\
&=\nonumber
\sum_{\langle\mathcal{P}, \mathcal{S}, \mathcal{G}\rangle\in \mathcal{D}}
\indicator{ \exists p\in \mathcal{P}:\;\tau\leq \text{score}_\theta(p|\mathcal{P}, \mathcal{S})}\nonumber\\ 
&\text{ActualPositives}=\sum_{\langle\mathcal{P}, \mathcal{S}, \mathcal{G}\rangle\in \mathcal{D}}
\indicator{ \exists p\in \mathcal{P}:\;p\vdash_\textit{IO}\mathcal{G} }\nonumber
\end{align}
We  sweep possible values for $\tau$ to compute a precision-recall curve.
Generically, there is no `true' best trade-off between these desiderata.

\textbf{Pass rate.} 
The \emph{pass@k} metric~\citep{austin2021program, codex} measures the probability of $k$ samples from $P^\text{LM}(\cdot |\texttt{prompt})$ passing the ground-truth test cases, $\mathcal{G}$:
\begin{align}
\emph{pass@k}&= \E_{p_1, \cdots, p_k\sim P^\text{LM}(\cdot |\texttt{prompt})} \indicator{\exists p_i:\;p_i\vdash_\textit{IO}\mathcal{G}}
\end{align}
Note that \emph{pass@k} is proportional to ActualPositives (Eq.~\ref{eq:actualpositives}): The (fraction of) problems where there is at least one correct answer in the sampled programs.

It is also conventional to combine \emph{pass@k} with a scoring function that reranks the sampled programs.
This generalizes \emph{pass@k} to \emph{pass@k,n}, which measures the probability that, after generating $n$ candidate programs, a correct one is in the top-$k$ under our scoring function:
\begin{align}
&\emph{pass@k,n}= \nonumber\\
&\quad\quad\E_{\langle\mathcal{P}, \mathcal{T}, \mathcal{G}\rangle\sim \mathcal{D}} \indicator{
\begin{array}{l}
\exists p\in 
\text{top-k}_{p'\in  \mathcal{P}} \text{score}_\theta(p'|\mathcal{P}, \mathcal{T}) \nonumber\\
\text{ where }p\models\mathcal{G} 
\end{array}
}
\nonumber
\end{align}



\section{Results}

We study our approach on two popular datasets while using Codex models~\citep{codex} of different sizes, seeking to answer the following research questions:
\begin{itemize}
    \item Is the classifier well-calibrated? If so, how trustworthy can we make the system (precision), and how much does that require sacrificing coverage (recall)?
    \item How can we use the synthesized specifications to act as human-interpretable explanations of the behavior of the programs constructed by the language model?
    \item How does our learned reranking impact raw rate of success (\emph{pass@k,n})?
\end{itemize}
We evaluate  on programming problems from the Mostly Basic Python Problems~(MBPP:~\cite{austin2021program}, sanitized version) and HumanEval datasets~\citep{codex}.
Each of these datasets contains natural language descriptions of programming problems, and holdout tests to judge program correctness.
An important difference between them is that HumanEval sometimes includes example input-outputs as part of the natural language description, while MBPP does not.
Having I/O examples in the problem description makes spec generation easier: some specs are given for free.
On the other hand, humans sometimes spontaneously mix natural language and examples~\citep{larc}.
Therefore, using both MBPP and HumanEval gives a more robust evaluation, but we note this qualitative difference between them.
We gives further experimental setup details, such as hyperparameters and example prompts in the Appendix.

\subsection{Calibration, Precision, and Recall} 
\begin{figure}[!h]

\addtolength{\tabcolsep}{-1pt}    
\begin{tabular}{rccccccc}\toprule
&\multicolumn{3}{c}{\underline{\phantom{ttt}HumanEval\phantom{ttt}}}&\phantom{tt}&\multicolumn{3}{c}{\underline{\phantom{ttttttt}MBPP\phantom{ttttttt}}}\\
&\multirow{2}{*}{AUC}&max&R@&&\multirow{2}{*}{AUC}&max&R@\\
&&F1&P=.9&&&F1&P=.9
\\\midrule
\rowcolor{Gray}\multicolumn{8}{c}{Davinci (Largest Codex model)}\\



full & \textbf{0.91}&\textbf{0.91}&\textbf{0.91} && \textbf{0.74}&\textbf{0.80}&\textbf{0.43}\\
IO only & 0.86&0.88&0.83 && 0.69&0.75&0.41\\
rels only & 0.66&0.73&0.27 && 0.69&0.77&0\\
random & 0.15&0.39&0 && 0.23&0.48&0\\

\midrule
\rowcolor{Gray}\multicolumn{8}{c}{Cushman (Smaller Codex model)}\\


full & \textbf{0.67}&\textbf{0.72}&0.28 && \textbf{0.57}&\textbf{0.65}&0.18\\
IO only & 0.66&0.71&\textbf{0.33} && 0.55&0.63&\textbf{0.25}\\
rels only & 0.34&0.44&0.14 && 0.51&0.61&0.14\\
random & 0.08&0.28&0 && 0.16&0.40&0\\

\midrule
\bottomrule
\end{tabular}
\addtolength{\tabcolsep}{1pt}    
\begin{subfigure}[bh]{0.6\textwidth}
\includegraphics[width =0.67 \textwidth]{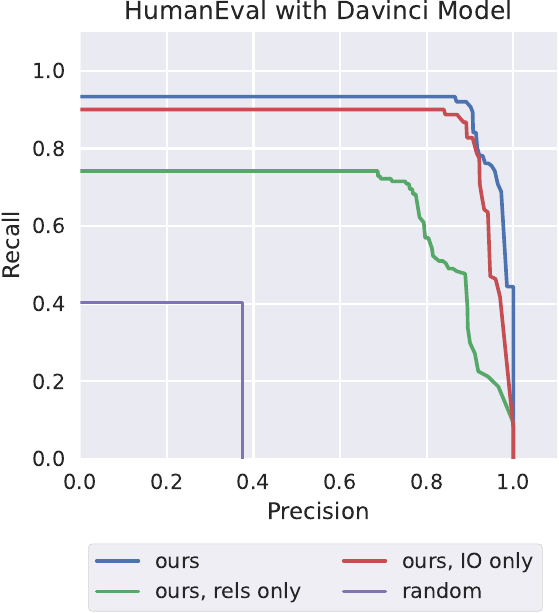}
\end{subfigure}
\caption{Top: Precision-Recall curves with $k=1$. 
Up: Statistics of these curves, measuring Area Under Curve (AUC), max F1 (harmonic mean of precision and recall), recall in the high-trust regime: R@P=.9 is recall when precision=90\%. The below figure gives further results.
}\label{fig:precisionrecall}
\end{figure}

Trustworthy systems should avoid predicting any programs at all when they cannot solve a problem.
Doing so increases precision, the fraction of outputs which are correct, but at the expense of recall, the fraction of solvable problems where we output a correct solution.
Fig.~\ref{fig:precisionrecall} illustrates how one can adjust this trade-off.
For example, we can achieve 100\% precision on HumanEval (\emph{zero} error rate), in exchange for dropping our recall from 93.4\% to 44.4\%.
Note this zero error rate does not come from our learned score function memorizing the data: we use cross validation to test each program using weights trained on other folds.
Less extreme tradeoffs are possible, such as 90\% precision in exchange for 90.7\% recall.

\begin{figure}[h]
\centering
\includegraphics[width=\columnwidth]{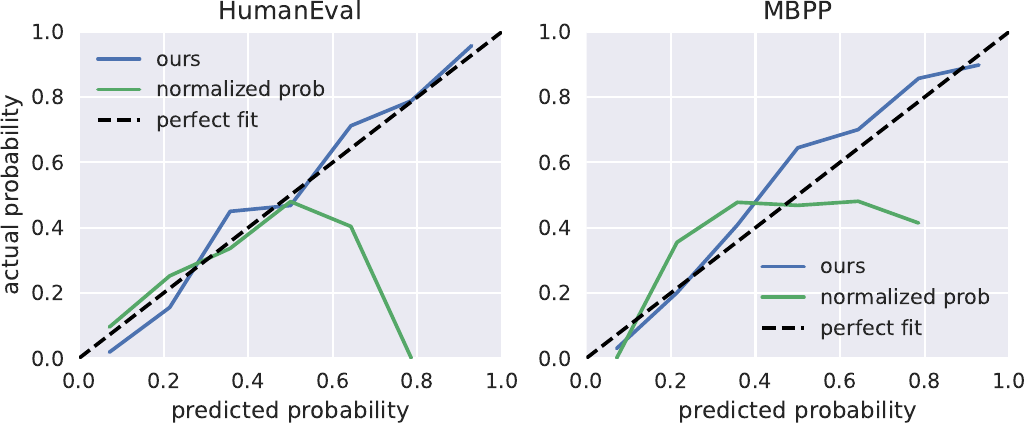}
\caption{Comparing the calibration of our probabilistic scoring function against raw Codex(Davinci) log likelihood.
Normalized probability, $Pr[\texttt{tokens}|\texttt{prompt}]^{\frac{1}{\texttt{tokens\_length}}}$, is the common metric used for scoring LLM samples~\citep{lin2022teaching,si2022prompting}.
}\label{fig:calibration}
\end{figure}

Striking favorable points on these tradeoffs is, in theory, a result of having a \emph{well-calibrated model}: whenever our probabilistic scoring function assigns probability $x$\% to a program being correct, then approximately $x$\% of the time, that program should actually be correct.
We confirm this calibration property in Fig.~\ref{fig:calibration}, also finding that raw log likelihoods from the language model are substantially less well-calibrated.
Calibration allows tuning the threshold $\tau$ to achieve a desired precision, because
the free parameter $\tau$  acts as a threshold on the probability of correctness needed to output a program.

Fig.~\ref{fig:precisionrecall} also shows that our full method typically performs best on precision/recall statistics in the most realistic setting, namely using the largest `Davinci' model.
Nonetheless, using just input-output specs, or just logical relations, can be effective on their own, as both outperform the random baseline which assigns a uniform probability to all programs sampled from the language model.

\subsubsection{A further application of calibration: Mixing language models of different sizes}

Large language models are expensive and energy-intensive, but often come in smaller, cheaper sizes.
In theory, predicting whether a language model can solve a problem should allow us to efficiently multiplex between a small cheap model and a large  powerful model by only delegating to the large model those problems which we predict cannot be solved by the smaller model.

\begin{figure}
    \centering
    \includegraphics[width=1\columnwidth]{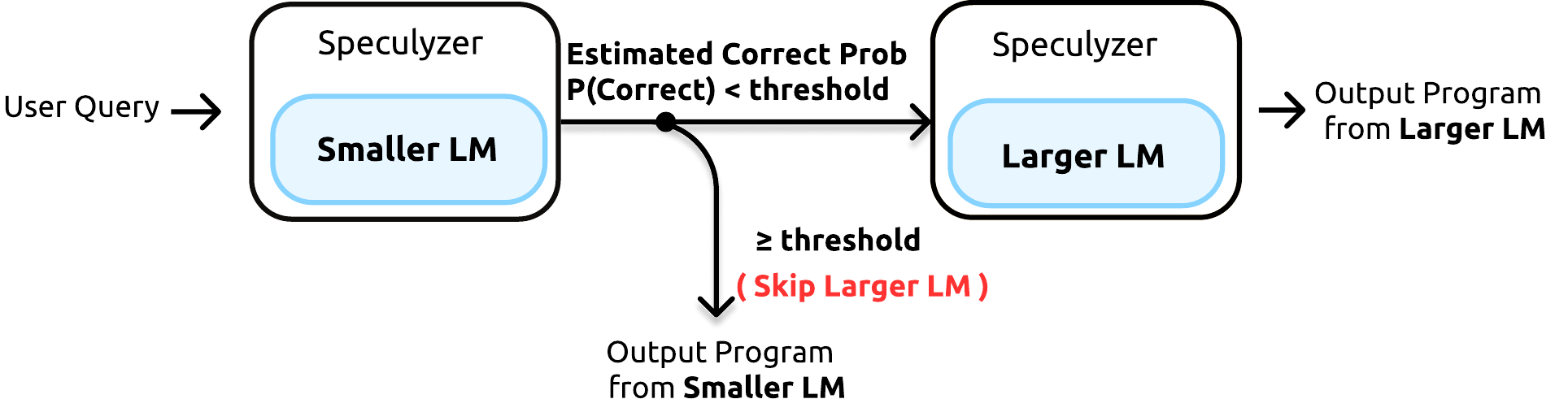}
    \caption{Model cascading with large and small models}
    \label{fig:model_mixing_diagram}
\end{figure}
\begin{figure}
    \centering
    \begin{subfigure}[b]{0.48\textwidth}
    \includegraphics[width=0.93\textwidth]{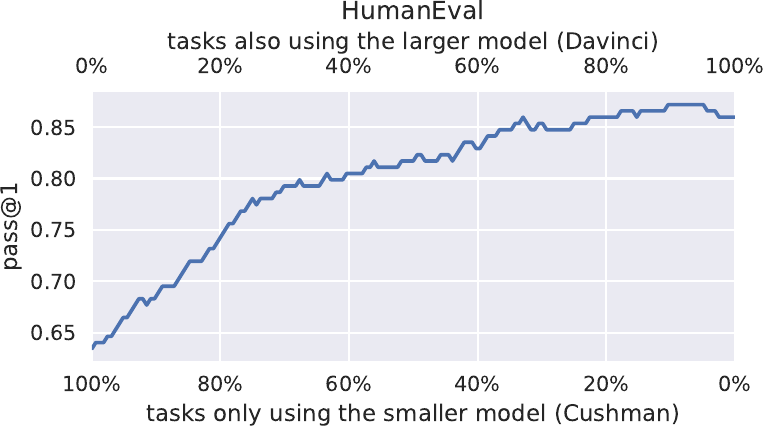}
\end{subfigure}
    \begin{subfigure}[b]{0.48\textwidth}
    \includegraphics[width=0.93\textwidth]{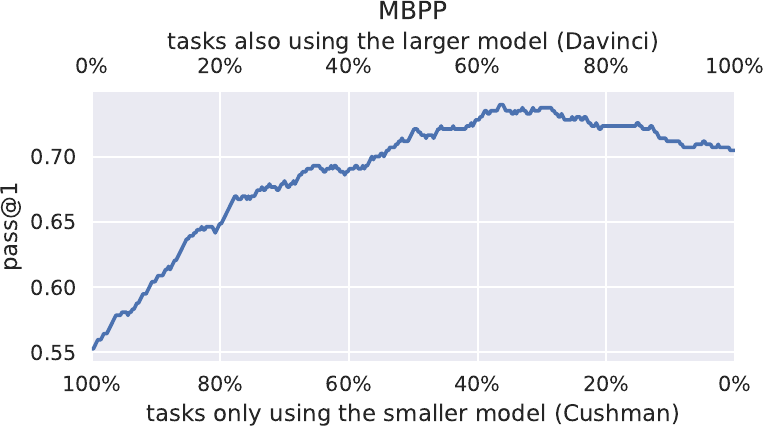}
    \end{subfigure}
    \caption{Mixing small cheap models and large powerful models by delegating to the large model only those problems that our approach thinks that the small model cannot solve.
    There are some MBPP problems where small model beats the big one. Our method multiplexes correctly for those.}
    \label{fig:model_mixing}
\end{figure}

We use our learned classifier to perform exactly this multiplexing, with Cushman-size Codex as the cheap model and Davinci-size as the powerful model as illustrate in Fig. ~\ref{fig:model_mixing_diagram}.
Using our probabilistic model to switch between them, we can approximately halve the number of queries to the large model while maintaining a similar number of solved programming problems (Fig.~\ref{fig:model_mixing}).
Contemporaneous work~\cite{chen2023frugalgpt} has also considered similar cascading of language models, but required bespoke algorithms to learn the multiplexing policy.
Here we show that a decent mixing of language model sizes can be accomplished `for free' when we train a well-calibrated classifier to predict whether a program is correct.


\subsection{Specs as explanations for program behavior}\label{sec:specificationranking}
No natural language program synthesizer will always produce correct programs:
Therefore, the system needs to explain what a synthesized program $p$ computes, so that the user can confidently accept or discard it.
\system{} does this by outputting a specification that is true about $p$,  while being maximally informative as to $p$'s behavior.
Below we formalize what it means to be `maximally informative', and describe a study with human participants that shows that they prefer the explanations generated by 
our approach, compared to the raw output of the LLM.

Whenever \system{} ranks $p^*\in \mathcal{P}$ as the best solution to a problem, it selects a spec $s^*\in \mathcal{S}$ to communicate the behavior of $p^*$.
The spec $s^*$ must be a true fact about $p^*$, so $p^*\vdash s^*$, but should also constrain the behavior of $p^*$.
For example, the specification $\forall x: p^*(x)=p^*(x)$ is vacuously true for any $p^*$, and so makes a poor explanation, despite holding for the ground-truth program.
As an example of a good explanation, if the program $p^*$  is incorrect, then the spec $s^*$ \emph{should also be incorrect}, in the sense that it is not true of the ground-truth program, despite holding for $p^*$.

\begin{figure}[!hbt]
\begin{subfigure}[b]{\columnwidth}
\center\includegraphics[width=\columnwidth]{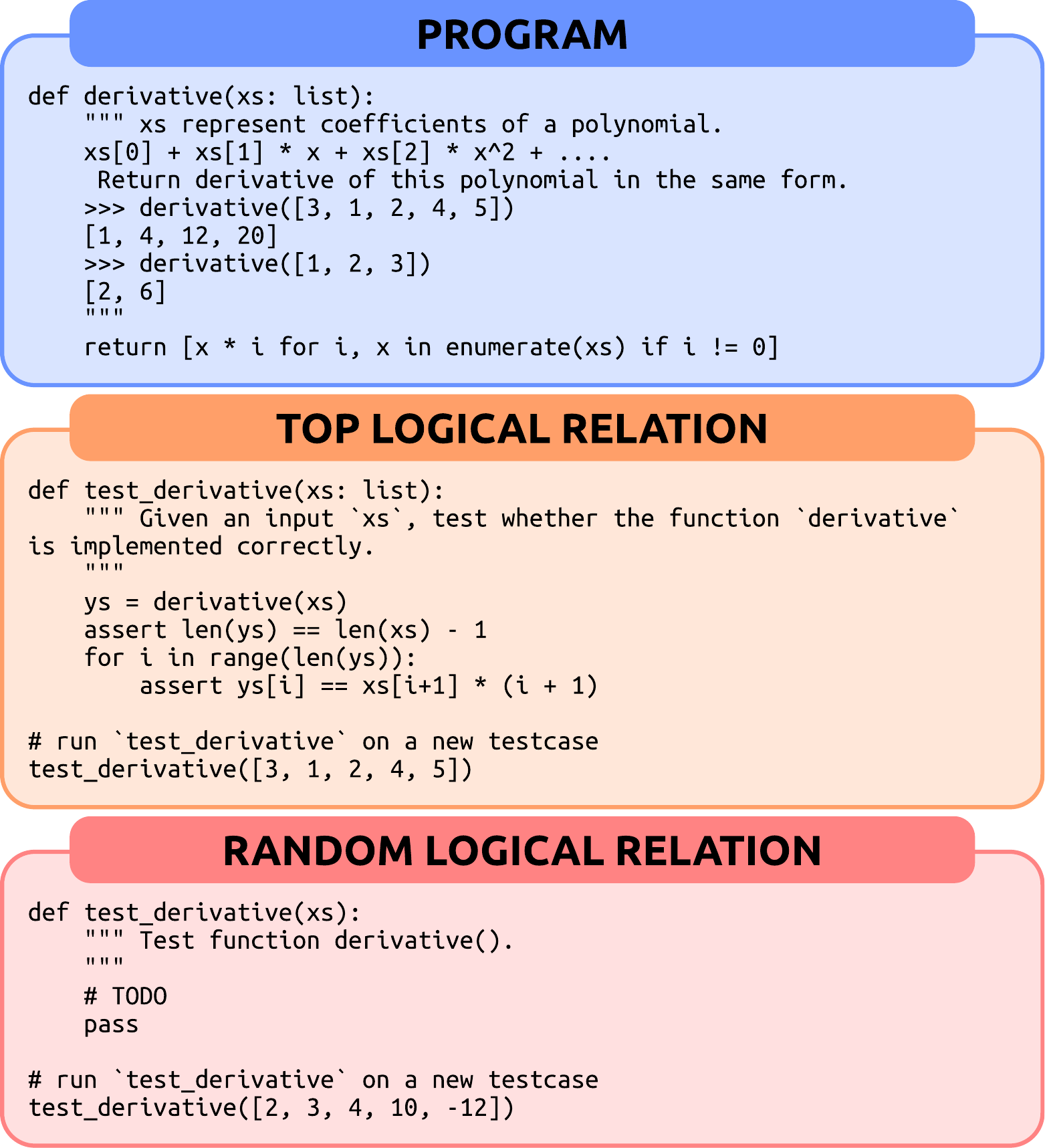}


\end{subfigure}
\caption{Program and specifications generated by language model and then reranked by our scoring function.
}\label{fig:userstudy}
\end{figure}

We formalize this as a rational-communication model of program synthesis~\citep{pu2020program}, which means first defining a joint probability distribution over programs and specifications:
$\probability{p,s}\propto \indicator{p\vdash s}\indicator{p\in \mathcal{P}}\indicator{s\in \mathcal{S}}$.
Then, we score each specification $s$ by the conditional probability of $p^*$ given $s$, i.e. $\probability{p^*|s}$.
Applying Bayes' Rule and simplifying, we find that this is equivalent to ranking specs by how few other programs satisfy them, i.e. their selectivity:
\begin{align}
s^*&=\argmax_{\substack{s\in \mathcal{S}}}\probability{p^*|s}=\argmin_{\substack{s\in \mathcal{S}\\p^*\vdash s}}\sum_{p\in \mathcal{P}}\indicator{p\vdash s}
\end{align}
To confirm that this is an effective method of scoring specifications, we recruited 22 human programmers, and ran an IRB-approved study where we showed them 8 programs synthesized by our system, together with the top-ranked spec, bottom ranked spec, and a random spec (Fig.~\ref{fig:userstudy}).
Every spec was a predicate that the program actually satisfied, and was the output of code-davinci-002.
Study participants rated each of the specifications on a 1-7 Likert scale, and were instructed to score how helpful the specification is in communicating whether the program is correct or incorrect.
Participants preferred the top-ranking specification over raw specs sampled from the LLM ($p<.05$, using a two-tailed $t$-test), and also dispreferred bottom-ranking specifications over the raw LLM output ($p<.05$, using the same statistical test).
However, the overall effect size was modest (Cohen's $d=0.14$), suggesting improving either the spec ranking heuristic or the underlying specs themselves (e.g., by fine-tuning the LLM) could be important in future work.
These results, however, clearly establish that the above heuristic is better than just the LLM on its own for generating interpretable explanations about program behavior that allow humans to decide whether a program is correct or not.





\subsection{\system{} for ranking programs}
So far we have shown how our system forms well-calibrated probability estimates of program correctness, which allows predicting whether a problem can be solved or not, in addition to inferring what the best program might be.
This is a harder, more general problem than reranking sampled programs to improve the accuracy of program synthesis.
How well then do our probability estimates serve as a ranking function, and does solving this harder problem sacrifice performance on the easier challenge of reranking sampled programs?
To answer this question, we measure \emph{pass@1,100} using our scoring function to rank sampled programs.
This means we sample 100 candidate programs, use our classifier to assign them probabilities, and then return the 1 program with the highest probability. 
We assess our system using repeated 5-fold cross validation and present the results in Figure \ref{fig:pass_figure}.
We can contrast with raw draws from the LLM (random), the state of the art approach to ranking samples from language models over code (CodeT), and
 a hypothetical Oracle that always chooses a correct program, if it exists.

Overall, \system{} achieves 85.7\% \emph{pass@1} on HumanEval and 70.5\% \emph{pass@1} on MBPP.
These are better than comparable published numbers for reranking methods~\cite{chen2022codet, inala2022fault, zhang2022coder, ni2023lever}.
This shows that we can achieve calibration via accurate probability estimates without sacrificing raw accuracy, and can even improve the state-of-the-art CodeT reranking method.

\begin{figure}[h]

\begin{tabular}{lccccc}
\toprule
Method&
Random & Oracle & Ours & CodeT*
\\\midrule
\rowcolor{Gray}\multicolumn{5}{c}{HumanEval}\\
cushman & 25.0&81.1&\textbf{63.3} \textpm 0.59&58.6\\
davinci &37.5&92.1&\textbf{85.7}  \textpm 0.73 & 74.8\\
\midrule
\rowcolor{Gray}\multicolumn{5}{c}{MBPP}\\
cushman &35.6&78.9&\textbf{55.6} \textpm 0.35 &55.4\\
davinci &44.6&84.8&\textbf{70.5} \textpm 0.24&67.7\\
\bottomrule
\end{tabular}
    \centering
    \caption{Pass@1,100: Calibration does not come at the expense of accuracy. CodeT results from \citet{chen2022codet}}
    \label{fig:pass_figure}

\end{figure}



\section{Contributions and Outlook}





We have contributed a program synthesizer that learns to predict when it cannot solve a problem and  selects specs that communicate what each program does.
We intend for these to increase the trust and safety of neural program synthesis and to serve as a modest step toward program synthesizers that could better collaborate with software engineers.
These advances do not come at the expense of raw accuracy, and also facilitate mixing neural networks of different sizes. 

Our work has important limitations. 
As \system{} wraps around a large language model, it inherits some of their limitations, 
such as expensive sampling times. Running the synthesized code incurs additional cost and imposes security risks if that execution is not appropriately sandboxed.
Fundamentally, an approach like ours can never truly provide the same level of trust as a program synthesis starting from human-crafted formal-logic specifications; however, formal logic is less accessible than natural language.

Many directions remain open. The idea of formal specifications as a liaison between programs and informal natural language opens up the possibility of using richer kinds of specs and verifiers, tapping years of effort from the programming languages community~\citep{d2008survey,baldoni2018survey}, at least if we can interface such formalisms with large language models.
Replacing program execution with sophisticated verification would also mitigate the aforementioned security concerns.
Another direction is integration with advanced HCI for program synthesis, such as~\cite{10.1145/3428227}, which develops powerful human interaction paradigms for program synthesis.

\bibliography{iclr2023_conference}

\input{neurips_2023_appendix}
\end{document}

%% file: math_commands.tex
\usepackage{amsmath,amsfonts,bm}

\def\eqref#1{equation~\ref{#1}}

\def\1{\bm{1}}

\usepackage{dsfont}
\newcommand{\indicator}[1]{\mathds{1}\left[ #1 \right]}

\DeclareMathAlphabet{\mathsfit}{\encodingdefault}{\sfdefault}{m}{sl}
\SetMathAlphabet{\mathsfit}{bold}{\encodingdefault}{\sfdefault}{bx}{n}

\newcommand{\E}{\mathbb{E}}

\DeclareMathOperator*{\argmax}{arg\,max}
\DeclareMathOperator*{\argmin}{arg\,min}

%% file: neurips_2023_appendix.tex
\clearpage

\appendix

\section{Experimental setup}\label{sec:hyper}
\noindent\textbf{Sampling from language models.} We used Codex models to draw samples using a max-token size of $300$ for our generation of programs and IO tests for both HumanEval and MBPP. For logical relation tests, we use $768$ and $512$ for the Davinci and Cushman models respectively.
We used  \texttt{"\textbackslash{ndef}"}, \texttt{"\textbackslash{n\#}"}, \texttt{"\textbackslash{nclass}"}, \texttt{"\textbackslash{nif}"}, and \texttt{"\textbackslash{nprint}"} as stop tokens for our generation of programs and input-output test cases, and we used \texttt{"\textbackslash{n\# Problem \{number\}}"} as the stop token for our generation of the logical relations test cases where \texttt{{number}} is set according to the number of few-shot examples. 
We used zero-shot prompting for program and input-output test case generation, and few-shot prompting for the logical relations specifications generation.
We drew samples from these models using nucleus sampling with temperature $=0.8$ and TopP $=0.95$.

\noindent\textbf{Logistic regressor.} We used the lbfgs solver from scitkit-learn
We used repeated 5-fold nested cross-validation with regularization hyperparameter $C=1$.

\noindent\textbf{Reproduciblity statement}
\begin{itemize}
    \item Data: The HumanEval and MBPP dataset are from existing literature and are available through Azure cloud service. We will also release the samples to facilitate reproducibility.
    \item Code: We detailed our hyperparameter and also we will make the code public upon publication.
\end{itemize}

\section{Dataset Statistics}
Below we show representative dataset statistics for Davinci Codex with temperature 0.8 and topP=0.95.

 \begin{adjustbox}{max width=\columnwidth}
{\centering\begin{tabular}{ccccc}
\toprule
&\multicolumn{2}{c}{Input-Output}&\multicolumn{2}{c}{Logical Relations}\\
\cmidrule(lr){2-3} \cmidrule(lr){4-5}
&HumanEval & MBPP 
&HumanEval & MBPP\\
\midrule
cluster size (\# of programs) & 12.91 & 13.62& 12.60 & 14.90\\
stddev & 18.02  & 20.52  & 17.82 &21.38\\
\midrule
average \# of test cases per program & 348.99 & 334.67 & 100 & 100\\
stddev & 137.18 & 146.47 & 0 & 0\\
\midrule
\begin{tabular}{c}
\% of programs that\\satisfy at least one test
\end{tabular}& 84.02\% & 82.49\% & 88.73\% & 82.43\% \\
\bottomrule
\end{tabular}}
\end{adjustbox}

\section{User Study}
In the framework of an Institutional Review Board (IRB)-approved study, we recruited 22 users, primarily consisting of Computer Science students. They were all compensated at rates exceeding the minimum wage, ensuring fair remuneration.

For the study, each participant was presented with a set of eight programs. These programs were samples from code-davinci-002 and consider top-rank programs by our system. Alongside each program, we offered three specifications: the top-ranked spec, the bottom-ranked spec, and a spec chosen at random. When the random-rank spec has the same score as the top-ranked or bottom-ranked spec, we re-sampled again.

Participants were requested to rate the utility of each specification using a 1-7 Likert scale. The scoring was based on their judgment of how effectively each spec assisted in determining the correctness or incorrectness of the corresponding program.

\begin{figure}
    \centering
    \includegraphics[width=0.6\columnwidth]{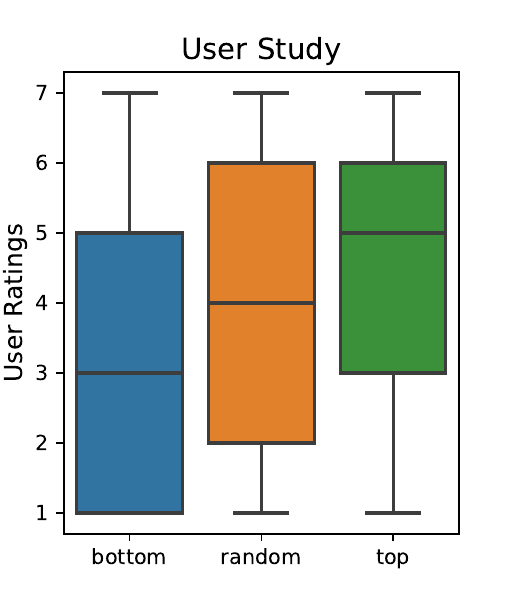}
    \caption{Boxplot representation of user ratings (1-7 Likert scale) for program specifications: top-ranked, random-ranked, and bottom-ranked. Participants prefer top ranked specifications over random ones, and random ones over bottom ranked specs.
   }
    \label{fig:study_result_boxplot}
\end{figure}

The results, illustrated in Fig. ~\ref{fig:study_result_boxplot}, establish that our heuristic is better than just the LLM on its own for generating interpretable explanations about program behavior that allow humans to decide whether a program is correct or not. The median values, around 5 for our approach, indicate a preference for the top-ranked specification over both the random and bottom ones.
Besides, in the first quantile, our approach outperforms the random specification.
In the third quantile, our heuristic achieves ratings of 6, surpassing the bottom-ranked specification.

\begin{figure}
    \centering
    \includegraphics[width=\columnwidth]{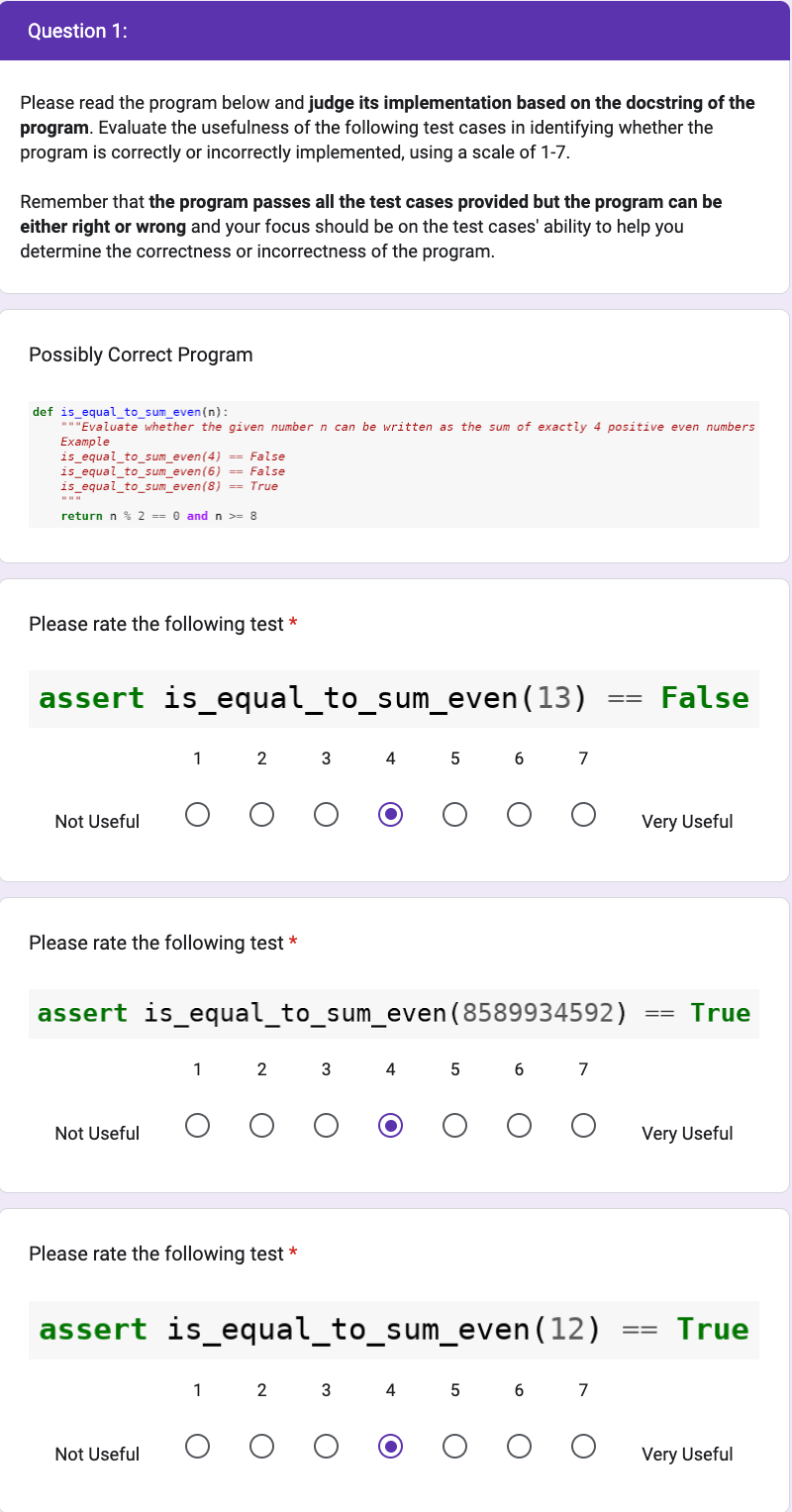}
    \caption{Screenshot of survey}
    \label{fig:survey}
\end{figure}

In Figure~\ref{fig:survey}, we provide a screenshot of the survey used for this study.

\begin{figure*}
    \begin{subfigure}[b]{\textwidth}
    \includegraphics[width=\textwidth]{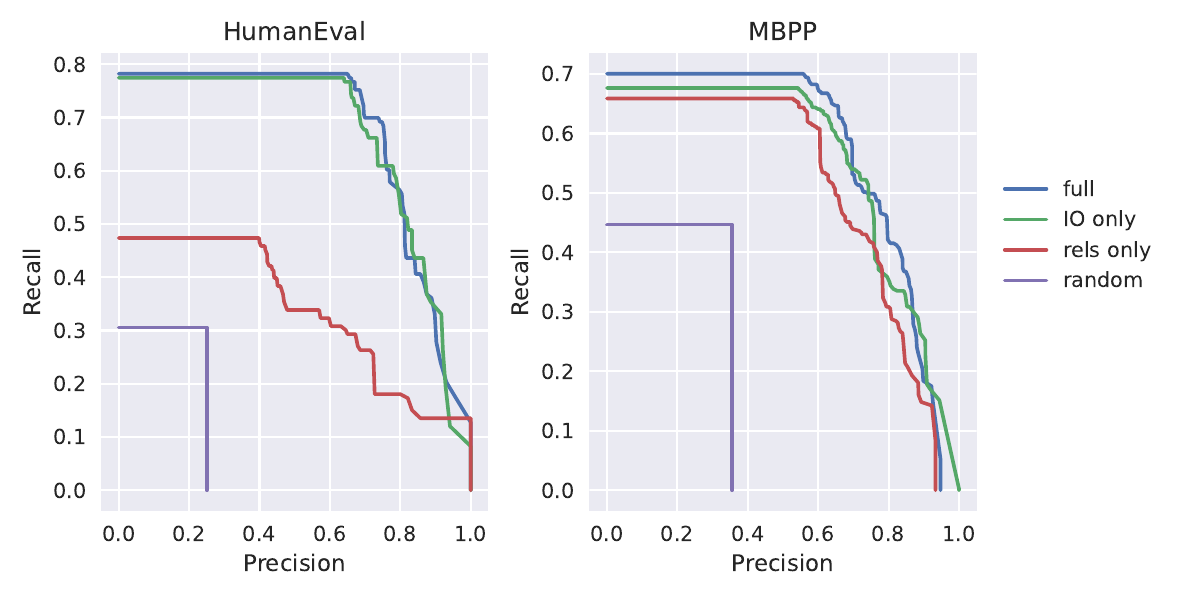}
    \caption{Model: cushman-code-001}
\end{subfigure}
    \begin{subfigure}[b]{\textwidth}
    \centering \includegraphics[width=\textwidth]{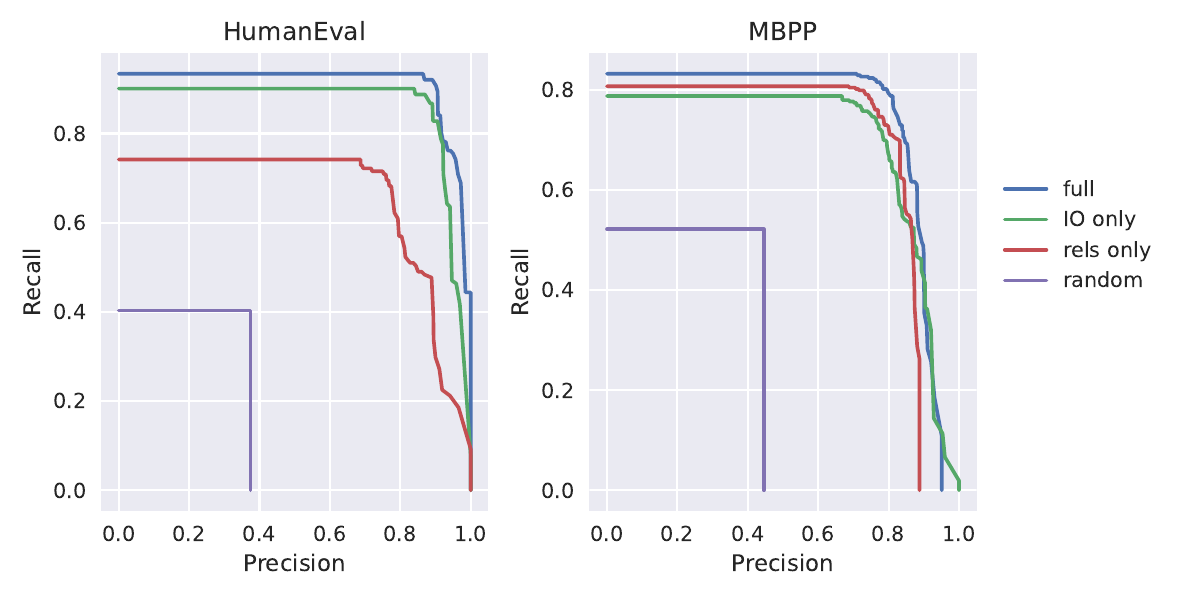}
    \caption{Model: davinci-code-002}
\end{subfigure} 
    \caption{Precision-Recall curves for each model and dataset}
\label{fig:all_precision_recall_curves}
\end{figure*}
\section{Precision-Recall Curves}
Figure ~\ref{fig:all_precision_recall_curves} presents four distinct precision-recall curves. Each of these curves corresponds to one of the four different settings delineated in the table, as shown in Figure \ref{fig:precisionrecall}. 

The illustrated curves provide a clear understanding of the relationship between precision and recall under various conditions. Interestingly, they demonstrate that high levels of precision can be achieved without substantially compromising the recall. This insight underscores the potential effectiveness of our approach in maintaining a balance between accurately identifying correct programs (precision) and still successfully retrieving correct programs in most cases(recall).

\section{Generalization across datasets}
Unlike recent heuristics for reranking solutions proposed by a large language model, our scheme involves learning real-valued parameters ($\theta$ in Eq.~(1).
To understand how learned parameters generalize across datasets,
 we compute the \emph{pass@1} rate and precision-recall stats for models trained on MBPP, but tested on HumanEval (and vice versa).
These statistics are similar by training on different datasets 
(Fig.~\ref{fig:generalization}), 
indicating generalization across similar, but not identical, data distributions.
\begin{figure*}[h]
\centering
\begin{tabular}{ccccc}
\toprule
test&\multicolumn{2}{c}{HumanEval}&\multicolumn{2}{c}{MBPP}\\
\cmidrule(lr){2-3} \cmidrule(lr){4-5}
train&HumanEval & MBPP 
&HumanEval & MBPP \\
\midrule
pass@1 & 0.86 & 0.76 & 0.68 & 0.70\\
AUC & 0.91 & 0.77 & 0.71 & 0.74 \\
max F1 &  0.91 & 0.81 & 0.76 & 0.80 \\
\bottomrule
\end{tabular}





\caption{Generalization when test/train data are drawn from the same corpus of problems, vs. drawn from different corpora.
}\label{fig:generalization}
\end{figure*}



\section{Example Zero-Shot Prompts for program generation}
For MBPP, to generate programs, we converted the natural language prompt to a function by adding in the prompt as a docstring for a function with the name of the function called in the ground-truth test cases. We used the HumanEval prompts as is.

Two examples of zero-shot prompts used for program generation are as follows:
\subsection{HumanEval}
First example:
\input{Prompts/he_program_prompt}

Second example:
\input{Prompts/he_program_2}
\subsection{MBPP}

First example:
\input{Prompts/mbpp_program_prompt}

Second example:
\input{Prompts/mbpp_program_2}

\section{Example Zero-Shot Prompts for input-output generation}
We extracted input-output test cases by generating $n=100$ times per HumanEval/MBPP prompt, then extracting each distinct single-line test case from each generation. We do this because each generation may produce multiple test cases, and we aimed to test each program on a single test case. For our test case prompts, we used the prompts to generate programs from MBPP and HumanEval, and we added in a \texttt{pass \# To-do: implement} statement, a line with a comment asking Codex to
\texttt{\# Check if func\_name works} and another line to asking  Codex to
\texttt{assert func\_name(}.
\subsection{HumanEval}
First example:
\input{Prompts/he_testcase_prompt}

Second example:
\input{Prompts/he_testcase_2}
\subsection{MBPP}
First example:
\input{Prompts/mbpp_testcase_prompt}

Second example:
\input{Prompts/mbpp_testcase_2}

\section{Few-Shot Prompt for logical relations spec generation}
We use two-shot and five-shot examples prompting to guide the model to tests various kinds of properties for 2k-context-size Cushman model and 8k-context-size Davinci model respectively.
\subsection{HumanEval (Cushman)}
\label{humaneval_fewshot_prompt}
\input{Prompts/HumanEval_spec_prompt_cushman}
\subsection{HumanEval (Davinci)}
\input{Prompts/HumanEval_spec_prompt_davinci}
\subsection{MBPP (Cushman)}
\label{mbpp_fewshot_prompt}
\input{Prompts/MBPP_spec_prompt_cushman}
\subsection{MBPP (Davinci)}
\input{Prompts/MBPP_spec_prompt_davinci}

\section{Transformation of input problems to logical relations prompts}
\label{logical_transformation}
\input{Prompts/transformation}

\section{Examples of Top Spec versus Random Spec}\label{sec:appendixdump}

In the following pages, we introduce 8 example HumanEval problems along with program samples generated by the code-davinci-002 model. For each of these programs, we display our top-ranked specification, a randomly chosen specification, and a bottom-ranked specification. Among them, four illustrate relation specifications, while the remaining present input/output (IO) specifications.

\begin{figure}
    \begin{subfigure}[b]{\columnwidth}
    \includegraphics[width=\textwidth]{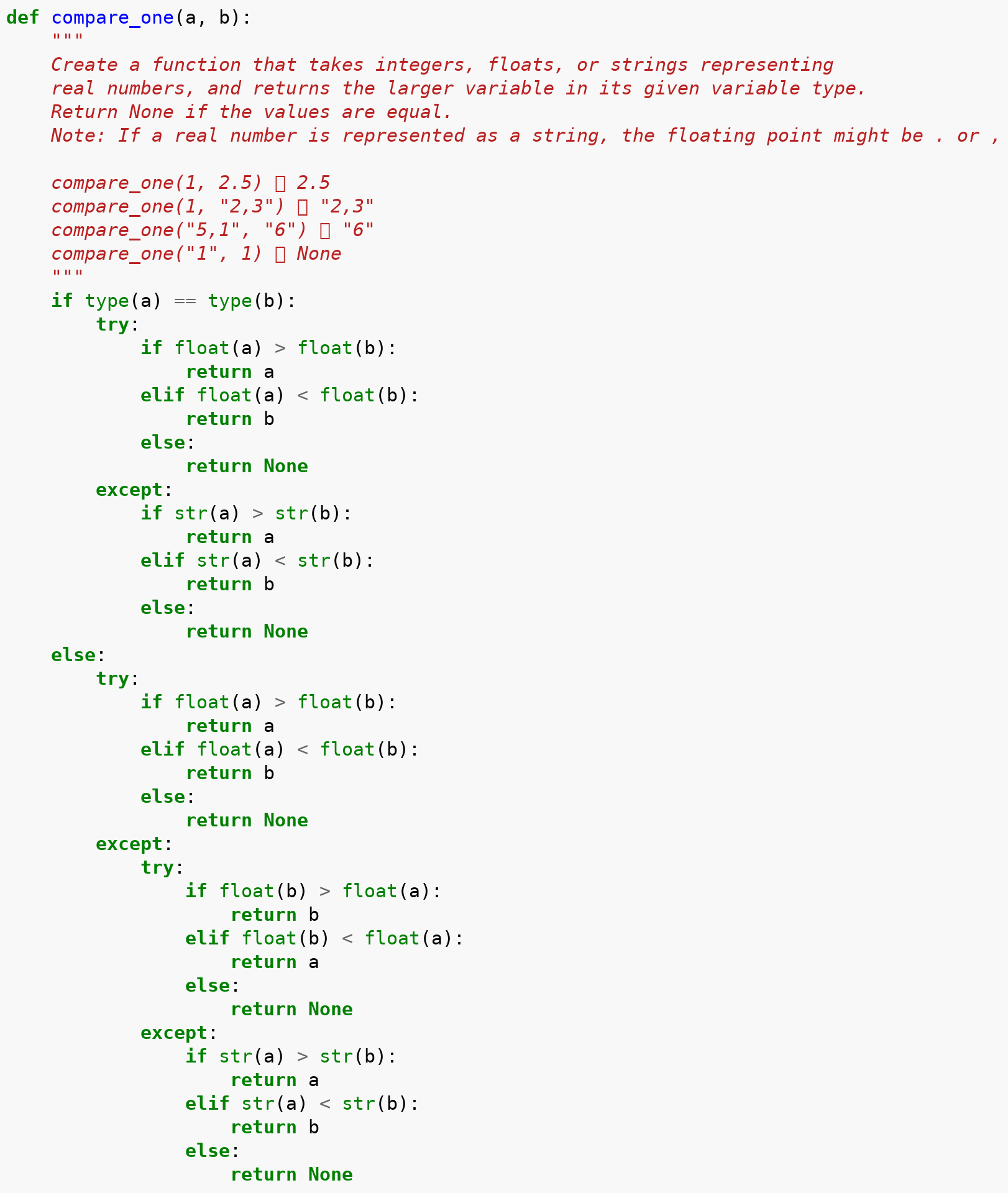}
    \caption{Program 1}
    \end{subfigure}
\end{figure}
\begin{figure}\ContinuedFloat
    \begin{subfigure}[b]{\columnwidth}
    \includegraphics[width=\textwidth]{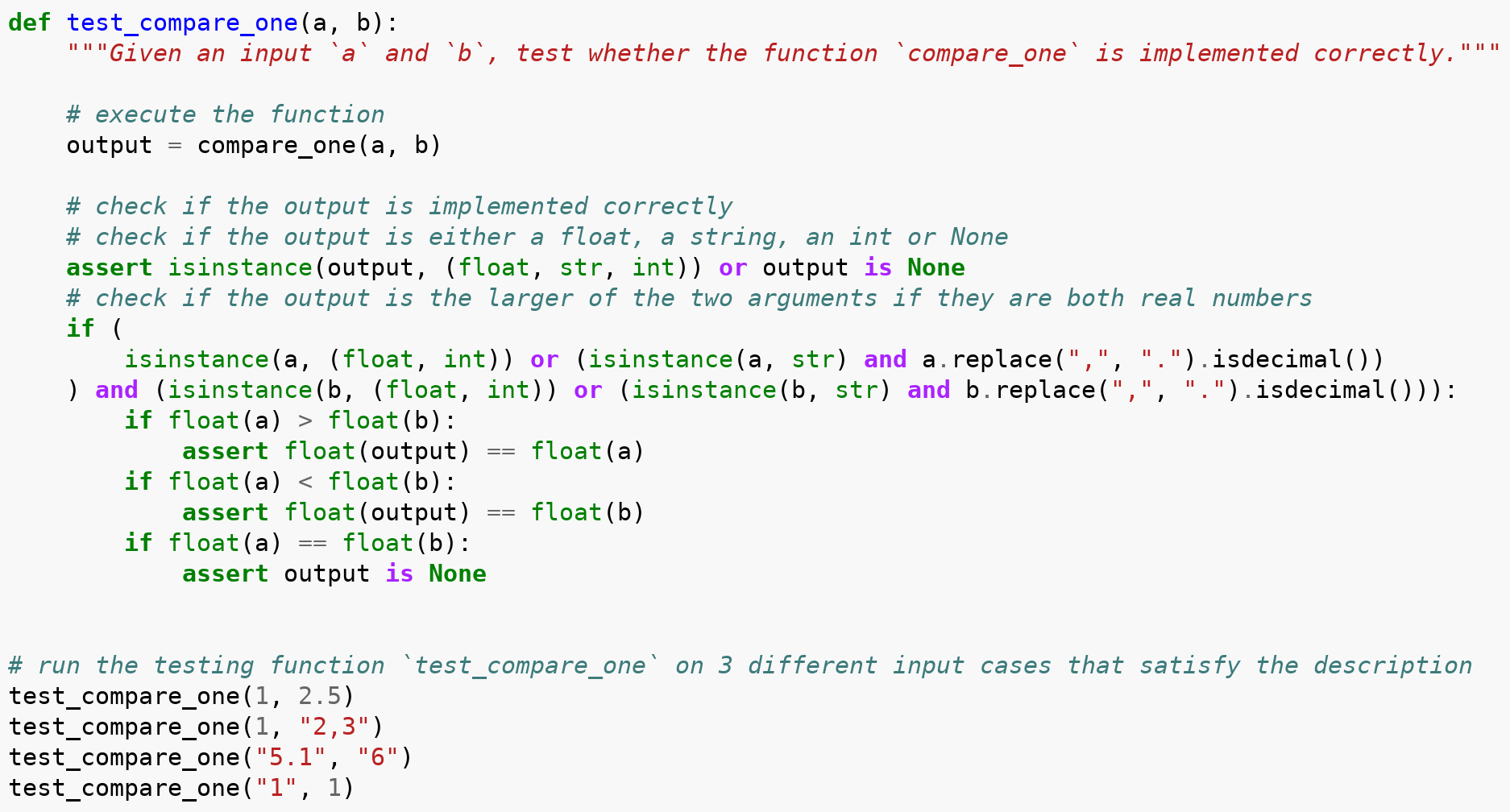}
    \caption{Top-ranked Relation Spec for Program 1}
    \end{subfigure}
    \begin{subfigure}[b]{\textwidth}
    \includegraphics[width=\textwidth]{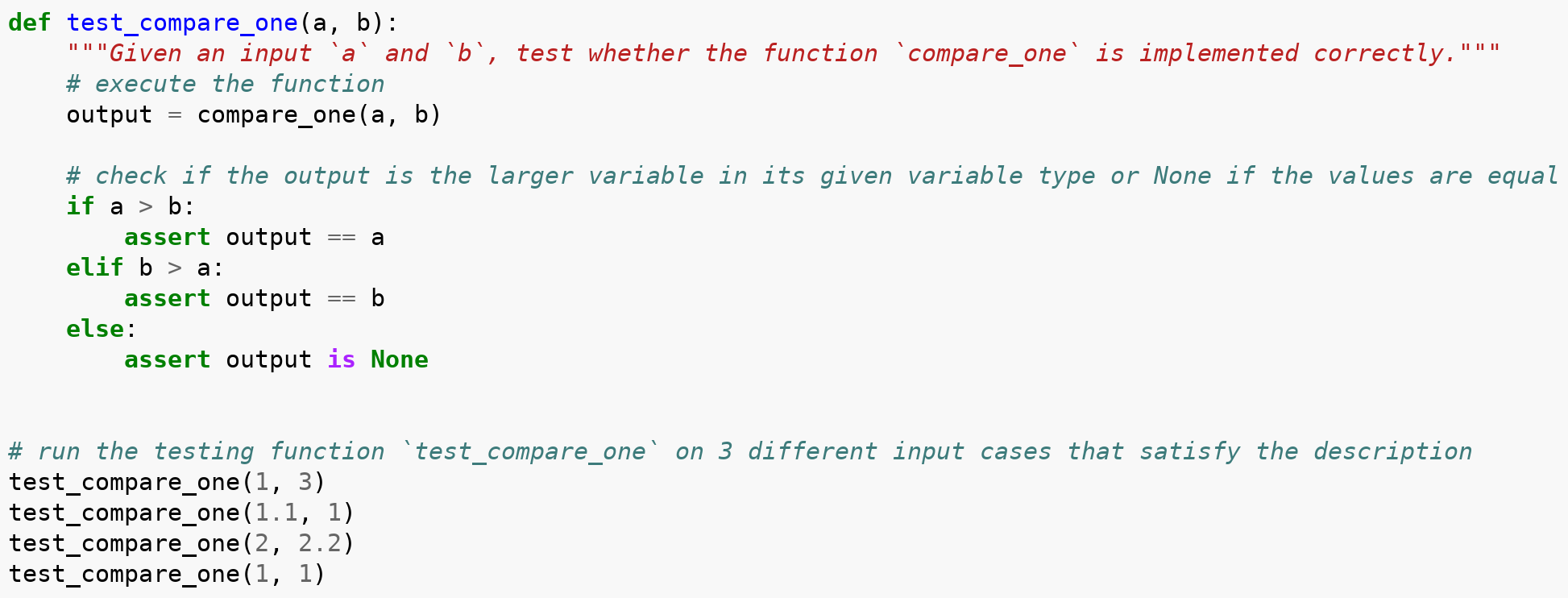}
    \caption{Random Relation Spec for Program 1}
    \end{subfigure}
    \begin{subfigure}[b]{\textwidth}
    \includegraphics[width=\textwidth]{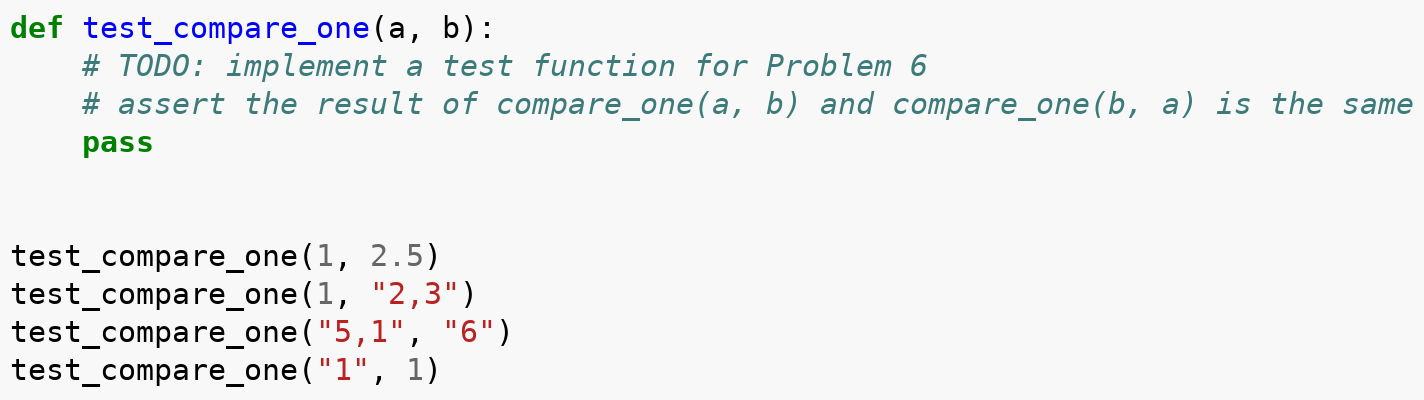}
    \caption{Bottom-Ranked Relation Spec for Program 1}
    \end{subfigure}
    \caption{Examples of top-ranked, random, bottom-ranked specifications}
\end{figure}
\begin{figure}
    \begin{subfigure}[b]{\columnwidth}
    \includegraphics[width=\textwidth]{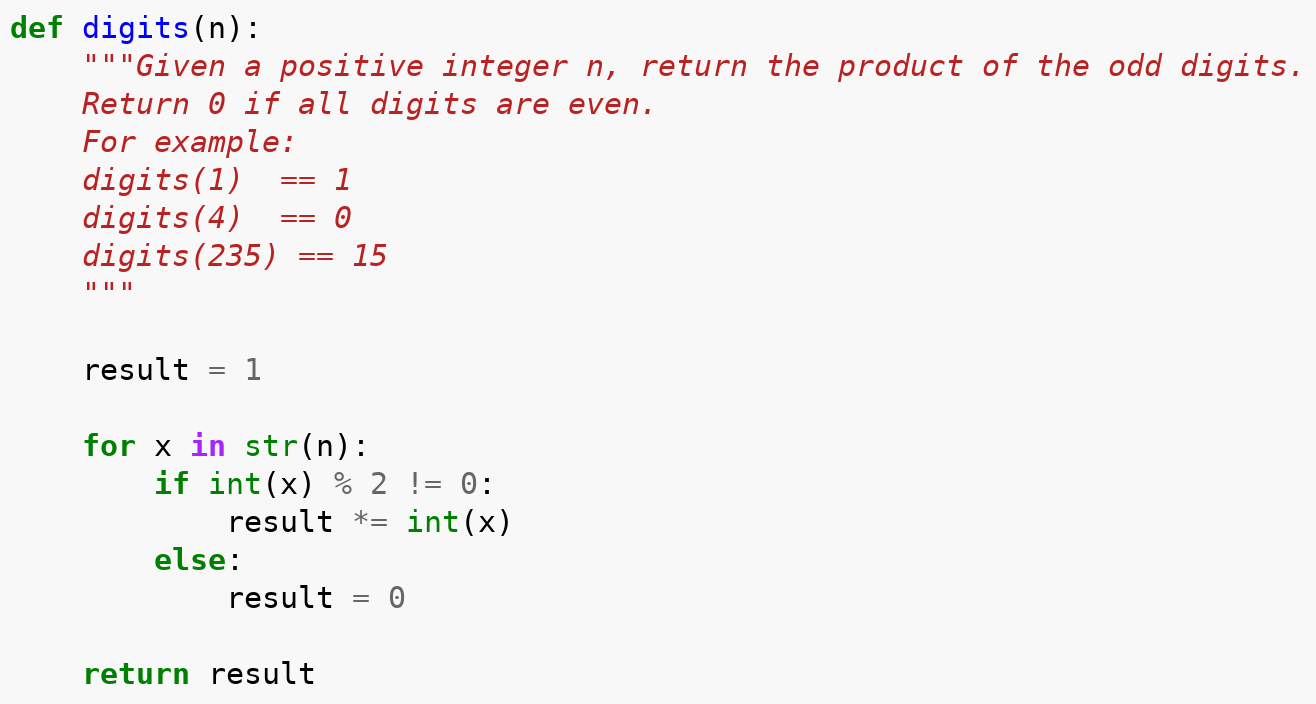}
    \caption{Program 2}
    \end{subfigure}
\end{figure}
\begin{figure}\ContinuedFloat
    \begin{subfigure}[b]{\columnwidth}
    \includegraphics[width=\textwidth]{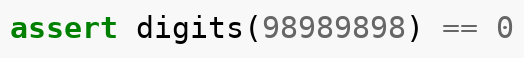}
    \caption{Top-ranked IO Spec for Program 2}
    \end{subfigure}
    \begin{subfigure}[b]{\columnwidth}
    \includegraphics[width=\textwidth]{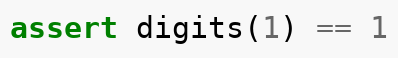}
    \caption{Random IO Spec for Program 2}
    \end{subfigure}
    \begin{subfigure}[b]{\columnwidth}
    \includegraphics[width=\textwidth]{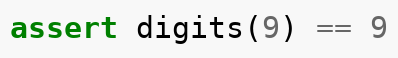}
    \caption{Bottom-Ranked IO Spec for Program 2}
    \end{subfigure}
    \caption{Examples of top-ranked, random, bottom-ranked specifications for Program 2}
\end{figure}
\begin{figure}
    \begin{subfigure}[b]{\columnwidth}
    \includegraphics[width=\textwidth]{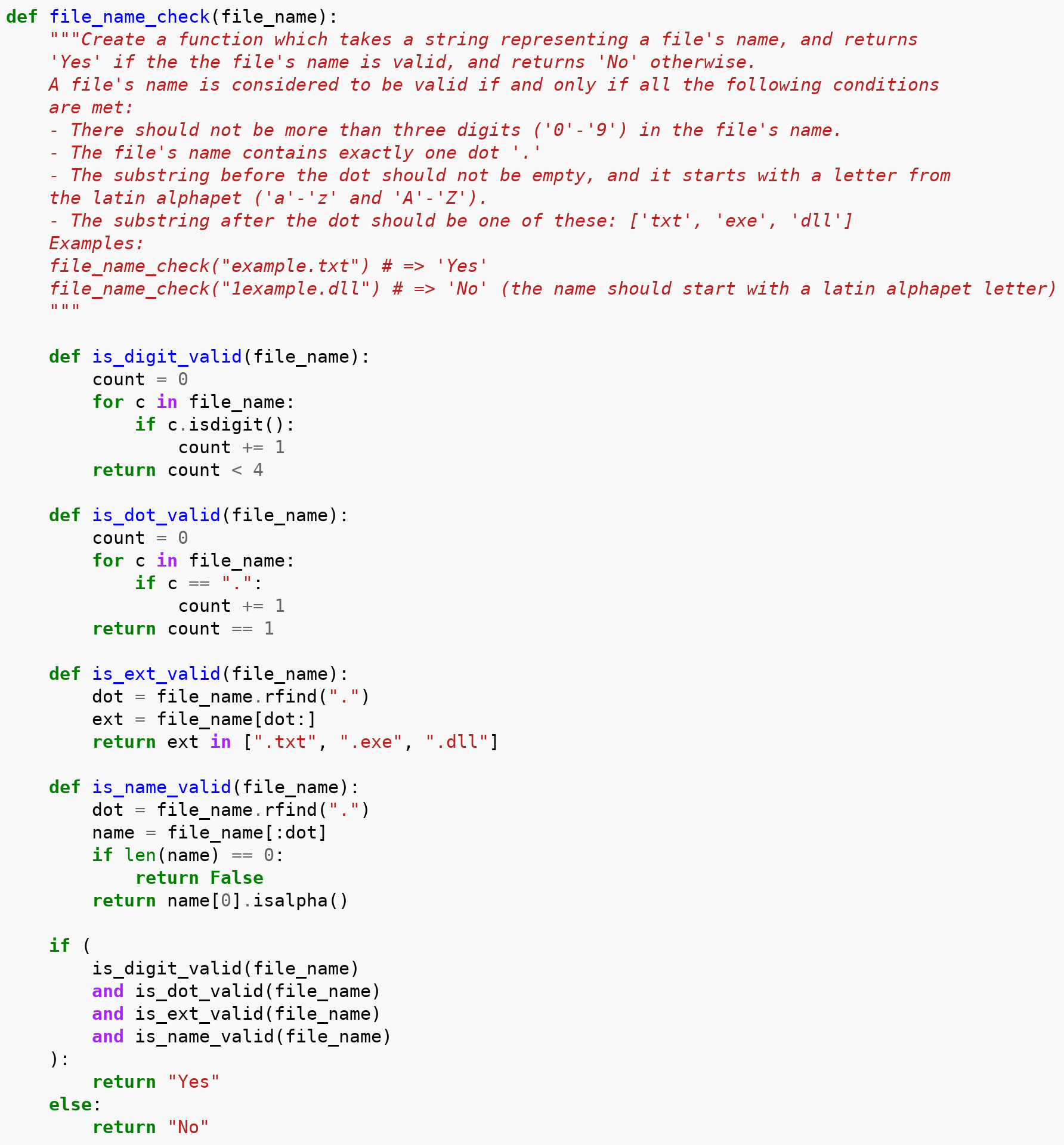}
    \caption{Program 3}
    \end{subfigure}
\end{figure}
\begin{figure}\ContinuedFloat
    \begin{subfigure}[b]{\columnwidth}
    \includegraphics[width=\textwidth]{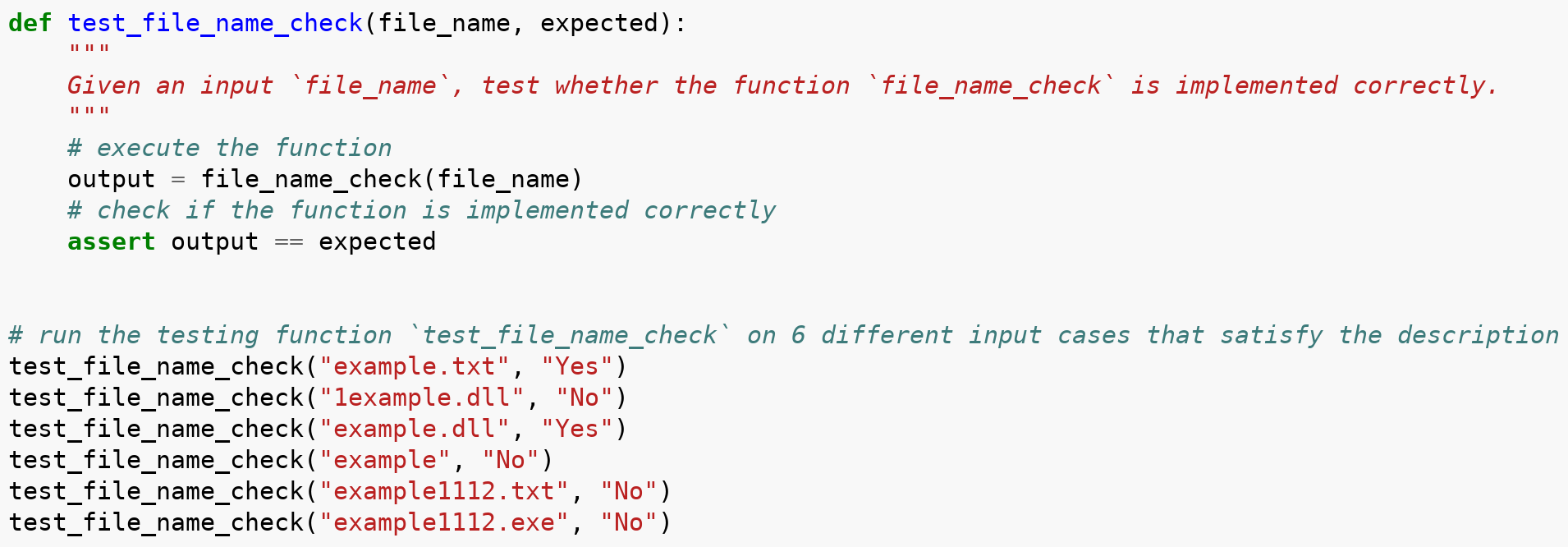}
    \caption{Top-ranked Relation Spec for Program 3}
    \end{subfigure}
    \begin{subfigure}[b]{\columnwidth}
    \includegraphics[width=\textwidth]{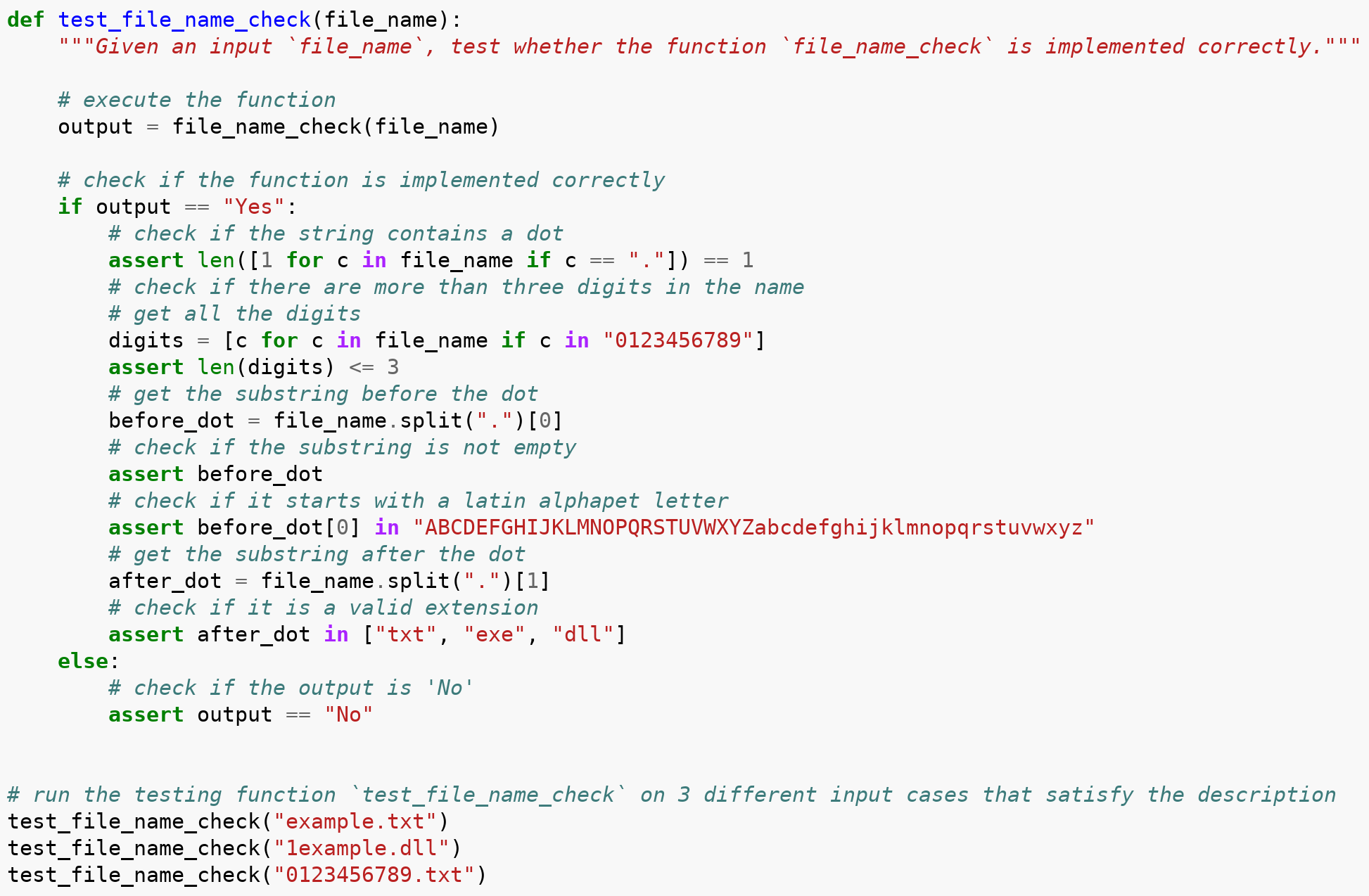}
    \caption{Random Relation Spec for Program 3}
    \end{subfigure}
    \begin{subfigure}[b]{\columnwidth}
    \includegraphics[width=\textwidth]{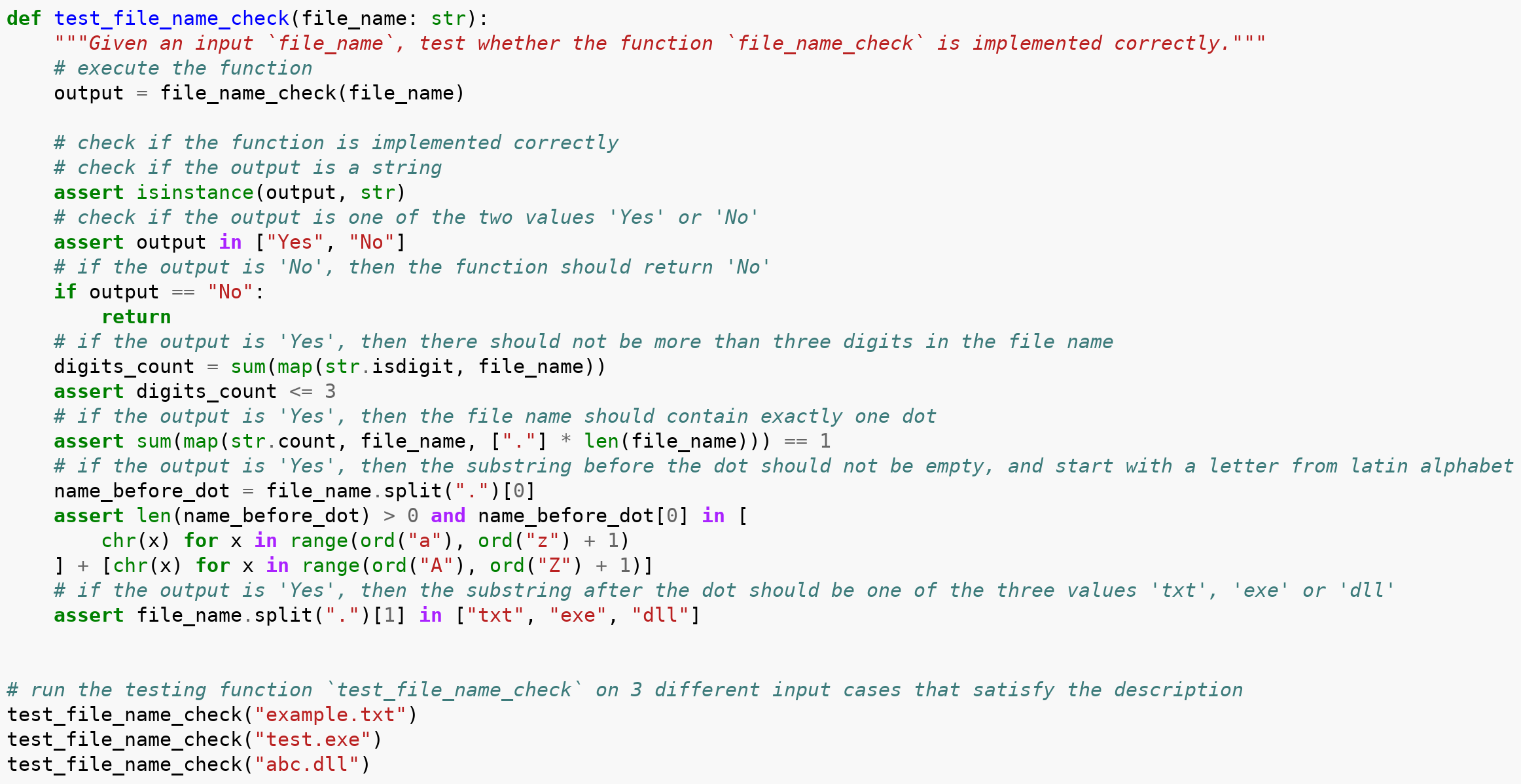}
    \caption{Bottom-Ranked Relation Spec for Program 3}
    \end{subfigure}
    \caption{Examples of top-ranked, random, bottom-ranked specifications for Program 3}
\end{figure}
\begin{figure}
    \begin{subfigure}[b]{\columnwidth}
    \includegraphics[width=\textwidth]{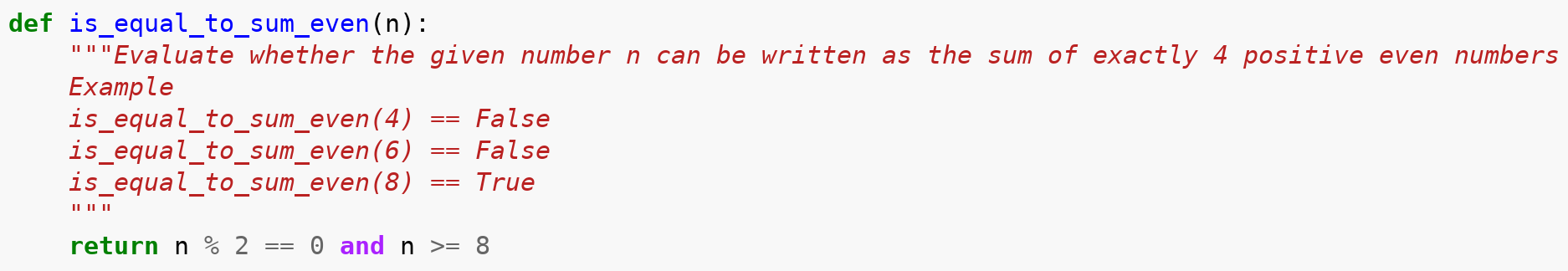}
    \caption{Program 4}
    \end{subfigure}
\end{figure}
\begin{figure}\ContinuedFloat
    \begin{subfigure}[b]{\columnwidth}
    \includegraphics[width=\textwidth]{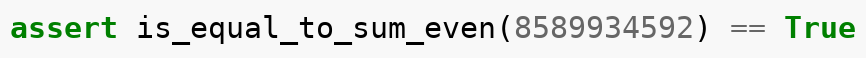}
    \caption{Top-Ranked IO Spec for Program 4}
    \end{subfigure}
    \begin{subfigure}[b]{\columnwidth}
    \includegraphics[width=\textwidth]{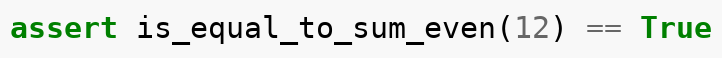}
    \caption{Random IO Spec for Program 4}
    \end{subfigure}
    \begin{subfigure}[b]{\columnwidth}
    \includegraphics[width=\textwidth]{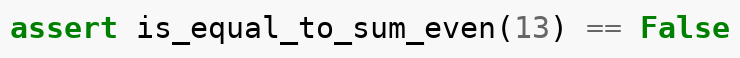}
    \caption{Bottom-Ranked IO Spec for Program 4}
    \end{subfigure}
    \caption{Examples of top-ranked, random, bottom-ranked specifications for Program 4}
\end{figure}
\begin{figure}
    \begin{subfigure}[b]{\columnwidth}
    \includegraphics[width=\textwidth]{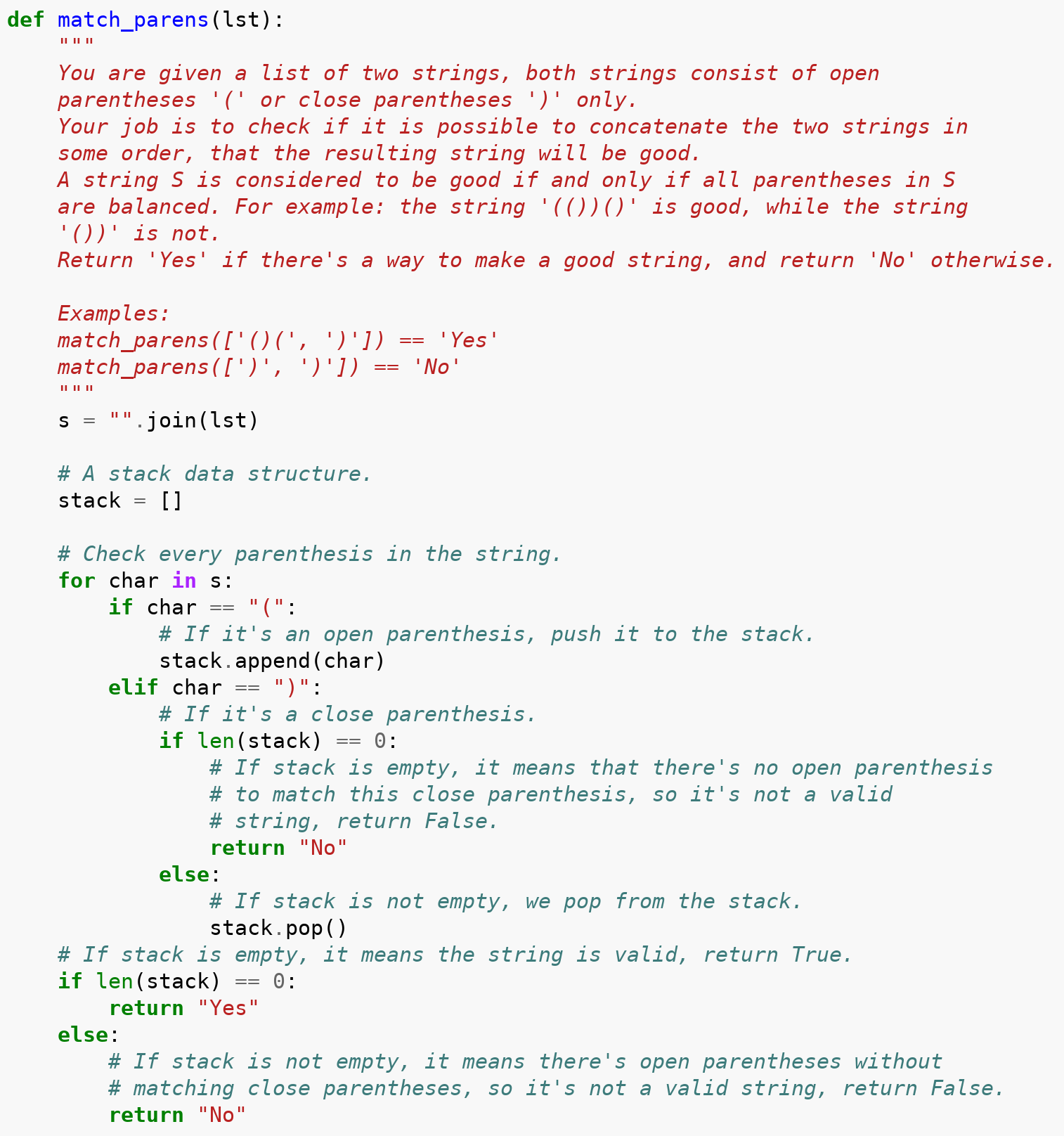}
    \caption{Program 5}
    \end{subfigure}
\end{figure}
\begin{figure}\ContinuedFloat
    \begin{subfigure}[b]{\columnwidth}
    \includegraphics[width=\textwidth]{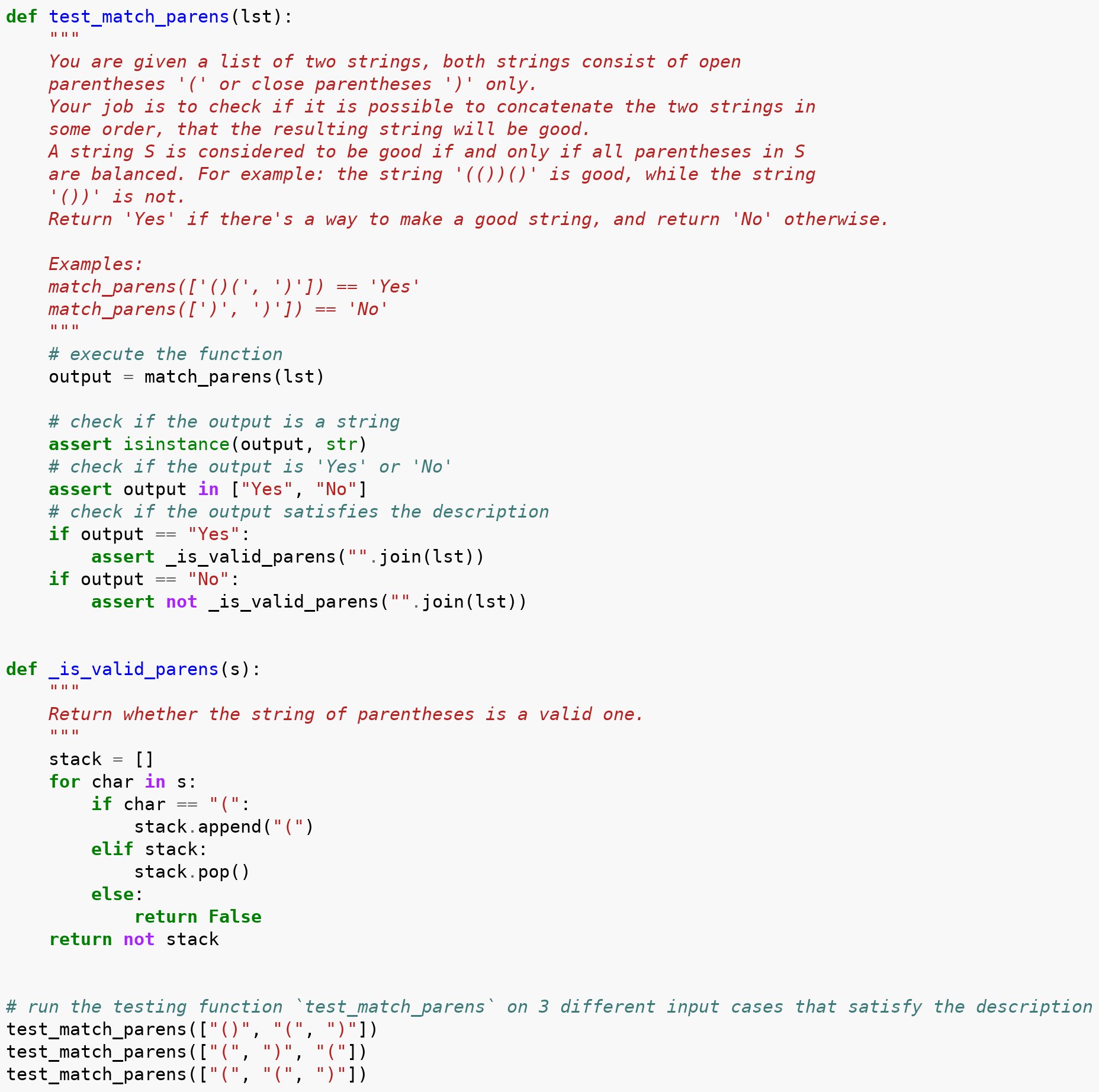}
    \caption{Top-ranked Relation Spec for Program 5}
    \end{subfigure}
    \begin{subfigure}[b]{\columnwidth}
    \includegraphics[width=\textwidth]{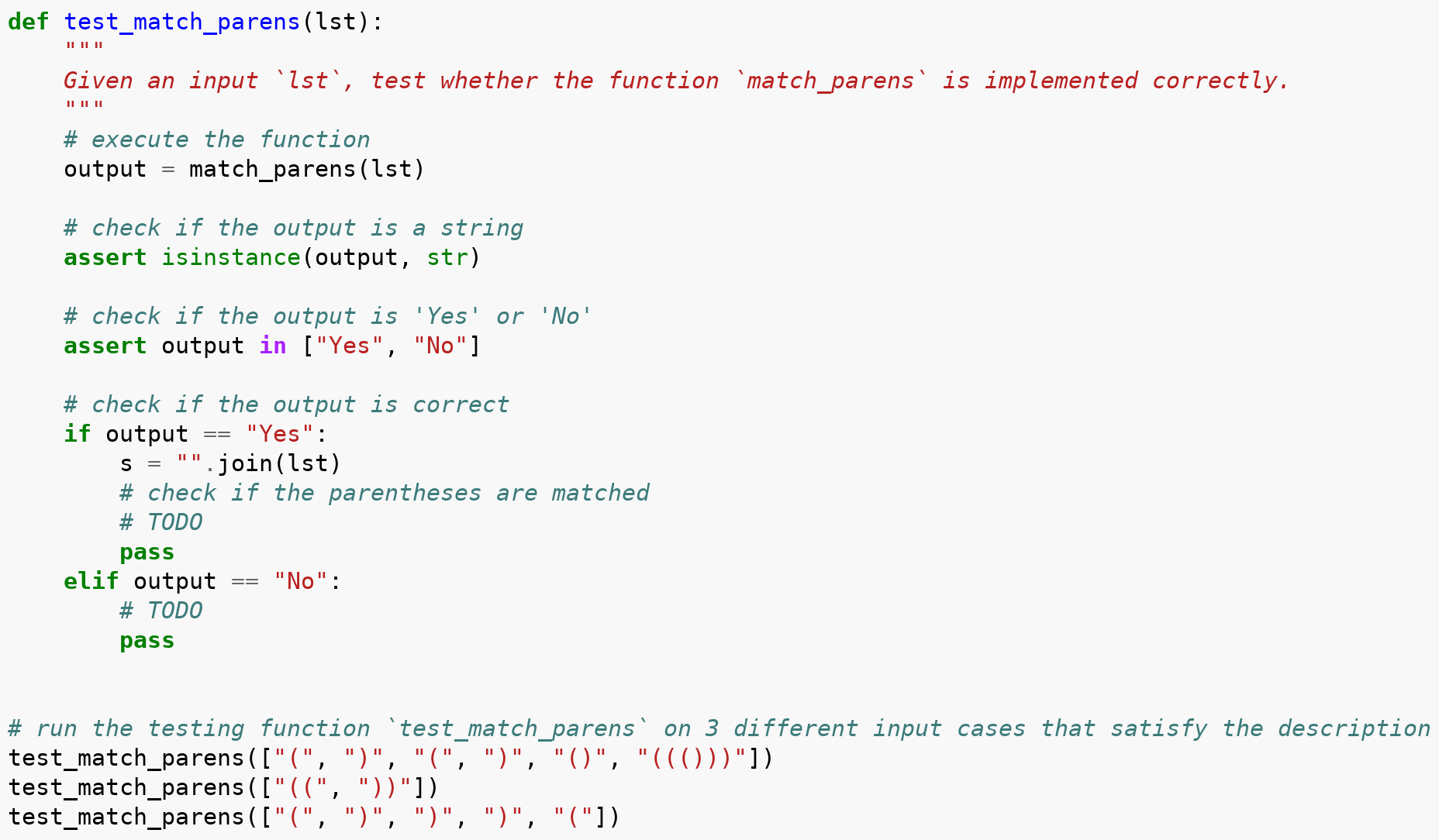}
    \caption{Random Relation Spec for Program 5}
    \end{subfigure}
    \end{figure}
    \begin{figure}\ContinuedFloat
    \begin{subfigure}[b]{\columnwidth}
    \includegraphics[width=\textwidth]{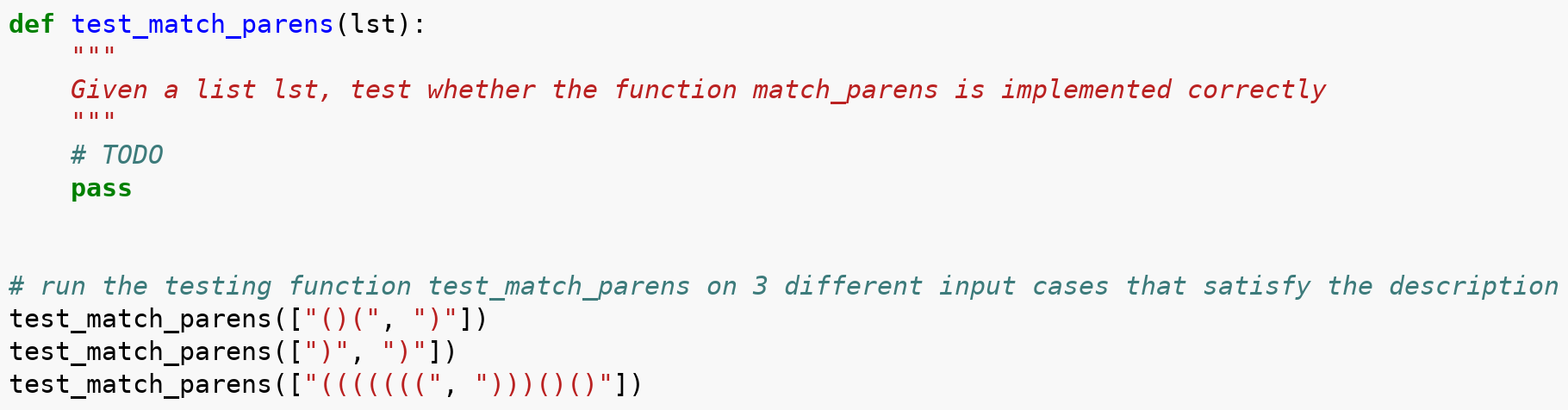}
    \caption{Bottom-Ranked Relation Spec for Program 5}
    \end{subfigure}
    \caption{Examples of top-ranked, random, bottom-ranked specifications for Program 5}
\end{figure}
\begin{figure}
    \begin{subfigure}[b]{\columnwidth}
    \includegraphics[width=\textwidth]{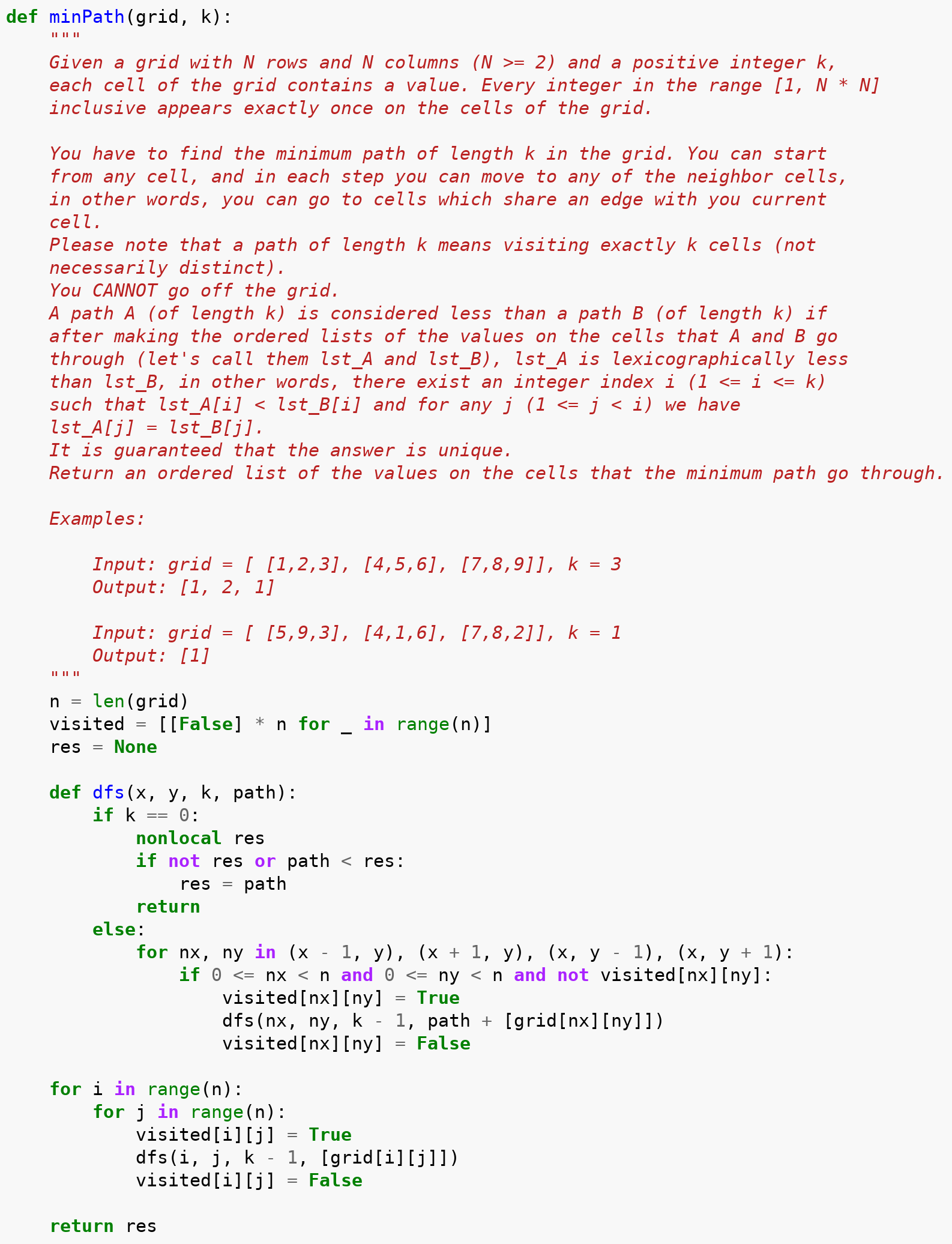}
    \caption{Program 6}
    \end{subfigure}
\end{figure}
\begin{figure}\ContinuedFloat
    \begin{subfigure}[b]{\columnwidth}
    \includegraphics[width=\textwidth]{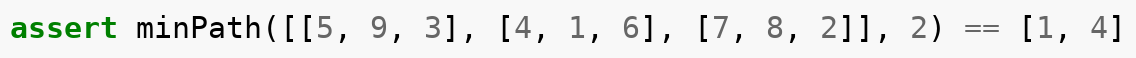}
    \caption{Top-ranked IO Spec for Program 6}
    \end{subfigure}
    \begin{subfigure}[b]{\columnwidth}
    \includegraphics[width=\textwidth]{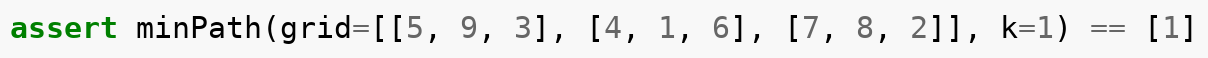}
    \caption{Random Spec IO for Program 6}
    \end{subfigure}
    \begin{subfigure}[b]{\columnwidth}
    \includegraphics[width=\textwidth]{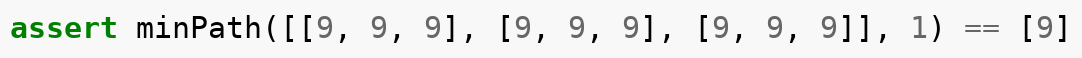}
    \caption{Bottom-Ranked IO Spec for Program 6}
    \end{subfigure}
    \caption{Examples of top-ranked, random, bottom-ranked specifications for Program 6}
\end{figure}
\begin{figure}
    \begin{subfigure}[b]{\columnwidth}
    \includegraphics[width=\textwidth]{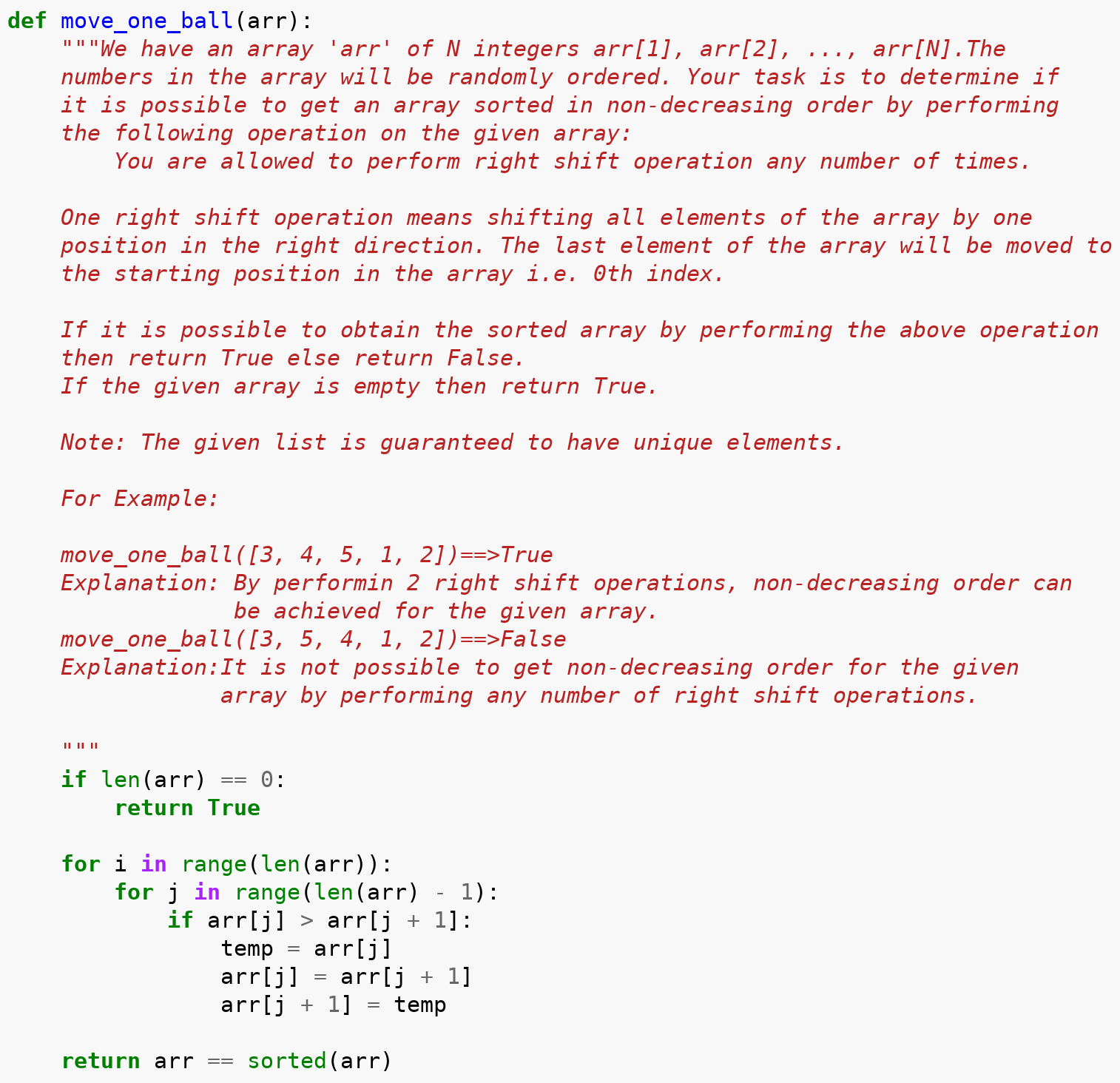}
    \caption{Program 7}
    \end{subfigure}
\end{figure}
\begin{figure}\ContinuedFloat
    \begin{subfigure}[b]{\columnwidth}
    \includegraphics[width=\textwidth]{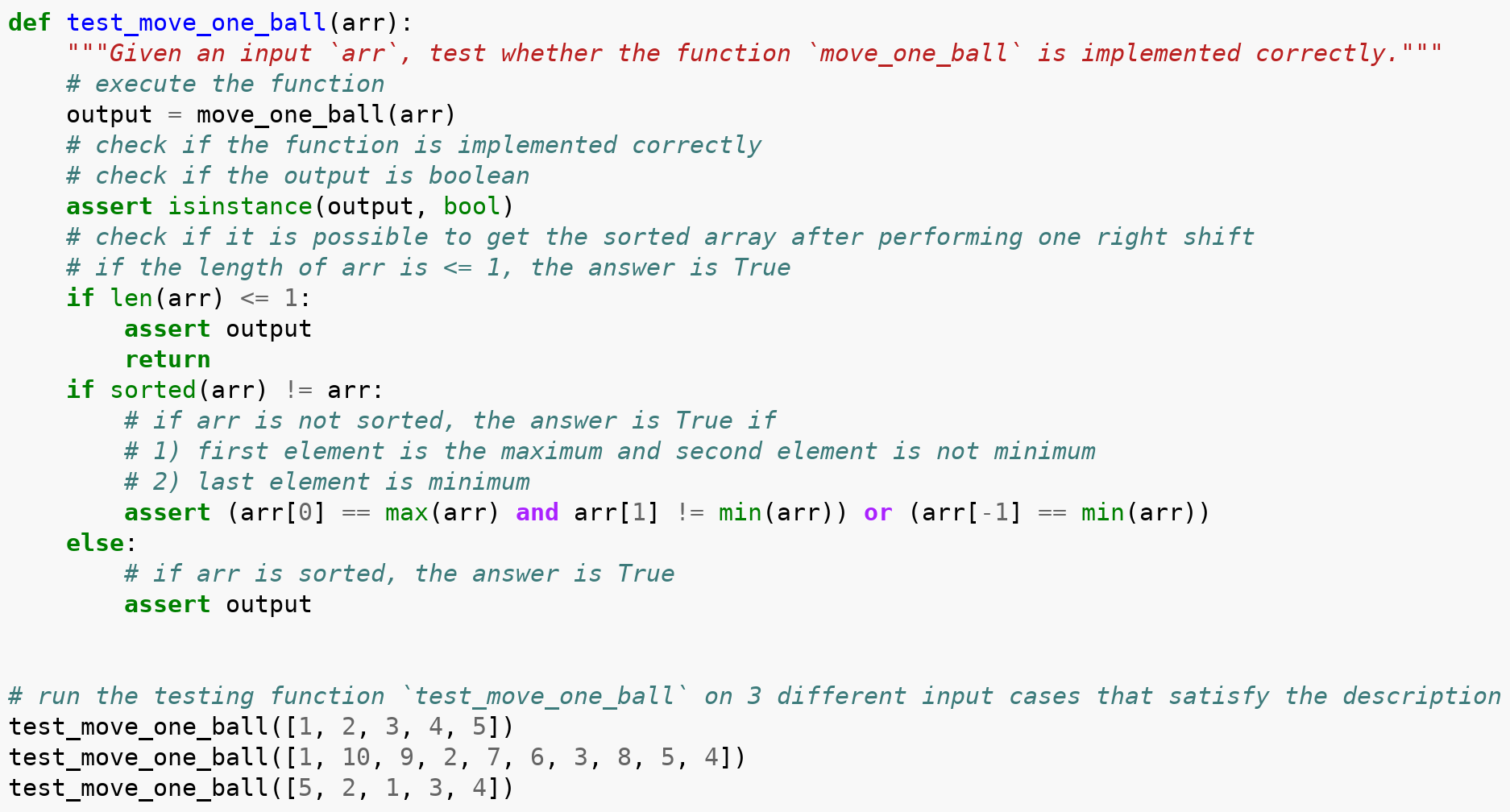}
    \caption{Top-ranked Relation Spec for Program 7}
    \end{subfigure}
    \begin{subfigure}[b]{\columnwidth}
    \includegraphics[width=\textwidth]{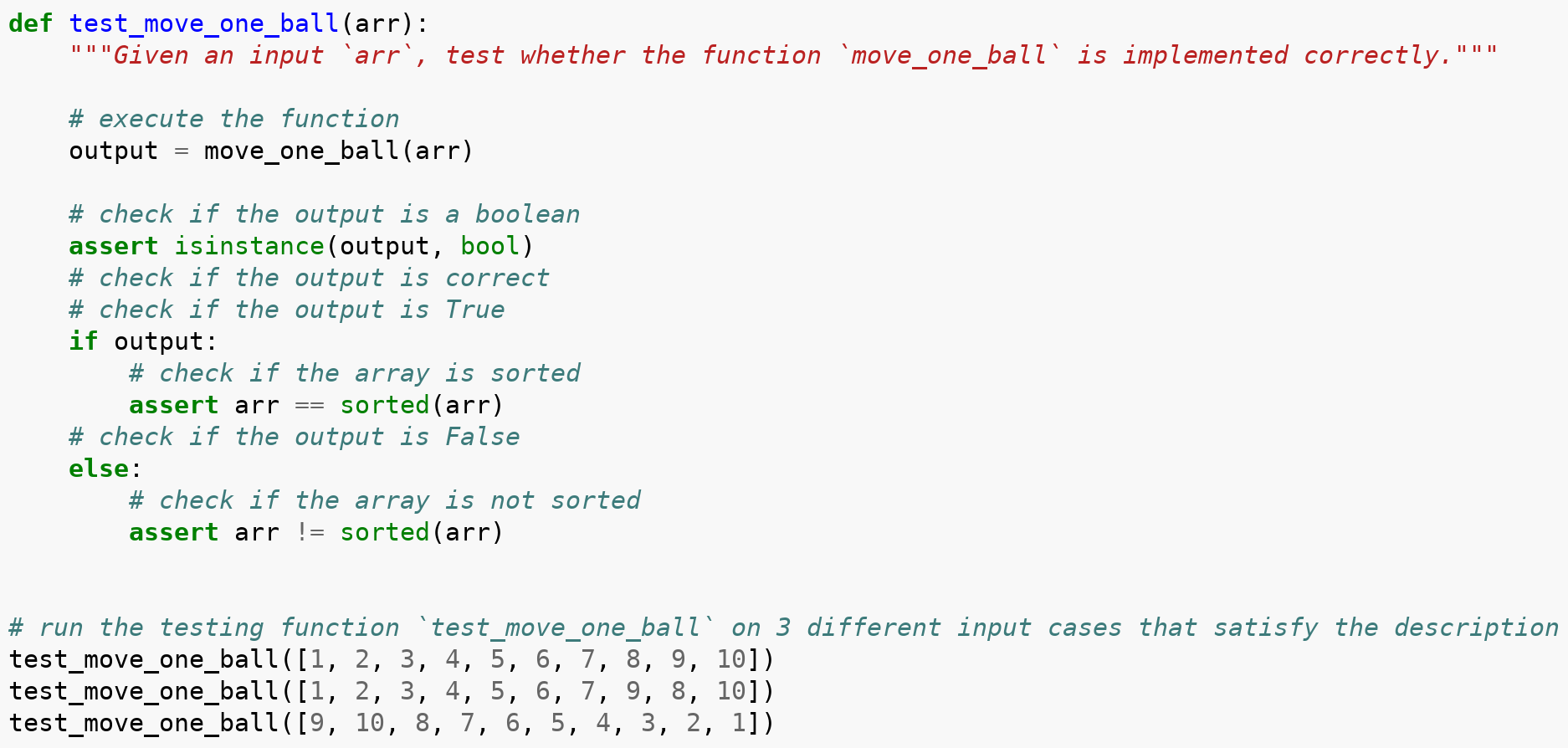}
    \caption{Random Relation Spec for Program 7}
    \end{subfigure}
    \begin{subfigure}[b]{\columnwidth}
    \includegraphics[width=\textwidth]{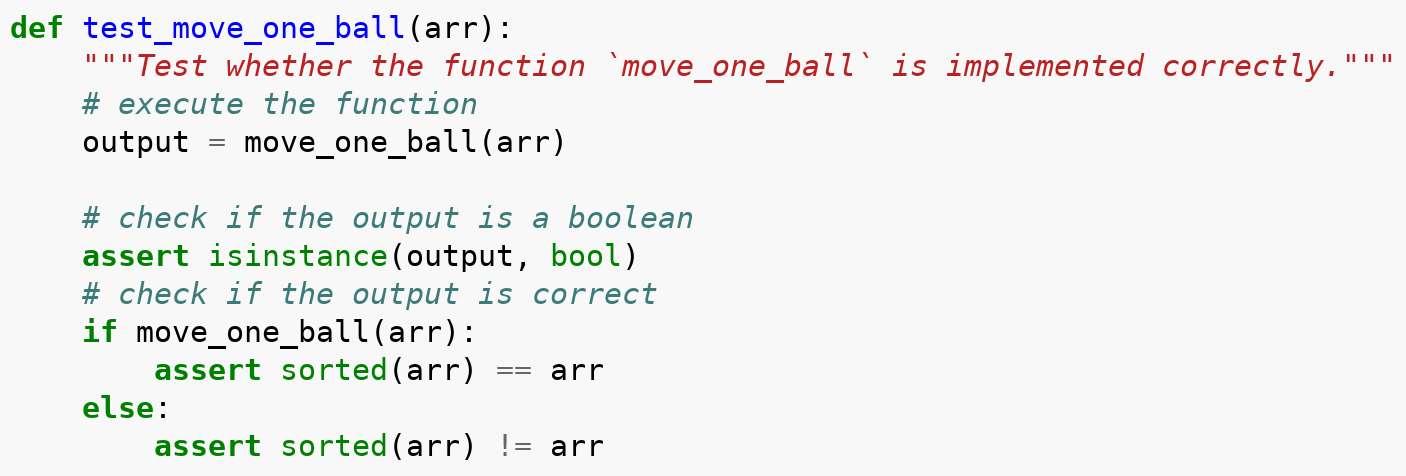}
    \caption{Bottom-Ranked Relation Spec for Program 7}
    \end{subfigure}
    \caption{Examples of top-ranked, random, bottom-ranked specifications for Program 7}
\end{figure}
\begin{figure}
    \begin{subfigure}[b]{\columnwidth}
    \includegraphics[width=\textwidth]{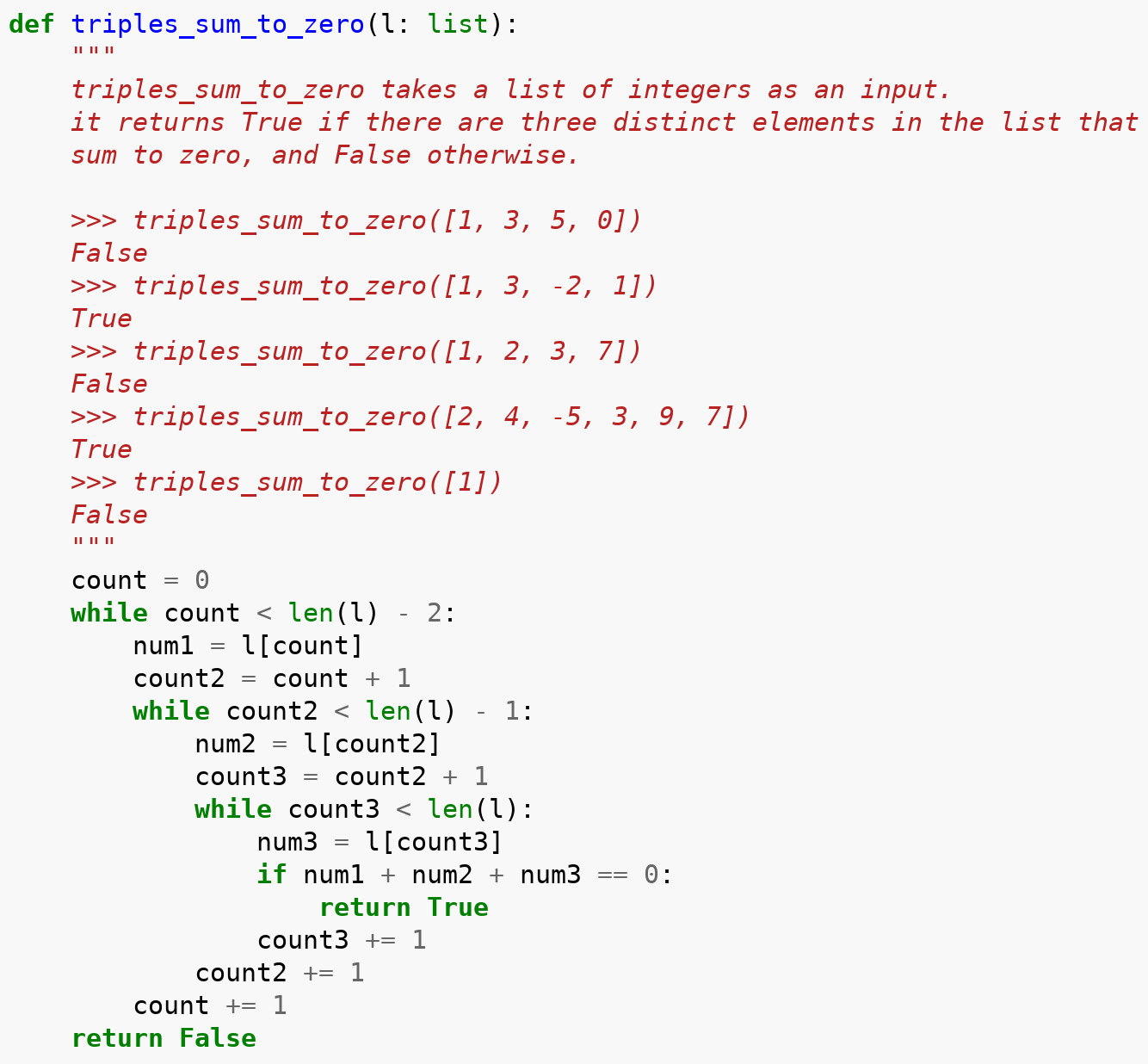}
    \caption{Program 8}
    \end{subfigure}
\end{figure}
\begin{figure}\ContinuedFloat
    \begin{subfigure}[b]{\columnwidth}
    \includegraphics[width=\textwidth]{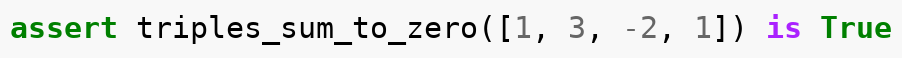}
    \caption{Top-ranked IO Spec for Program 8}
    \end{subfigure}
    \begin{subfigure}[b]{\columnwidth}
    \includegraphics[width=\textwidth]{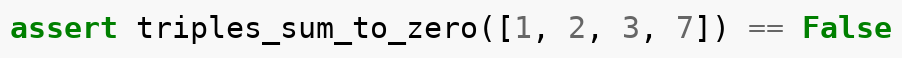}
    \caption{Random IO Spec for Program 8}
    \end{subfigure}
    \begin{subfigure}[b]{\columnwidth}
    \includegraphics[width=\textwidth]{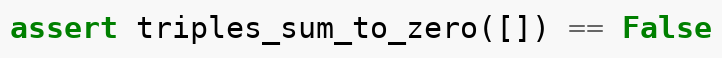}
    \caption{Bottom-Ranked IO Spec for Program 8}
    \end{subfigure}
    \caption{Examples of top-ranked, random, bottom-ranked specifications for Program 8}
\end{figure}




%% file: Prompts/he_program_prompt.tex
\begin{Verbatim}[breaklines=true, breakanywhere=true]
def is_happy(s):
    """You are given a string s.
    Your task is to check if the string is happy or not.
    A string is happy if its length is at least 3 and every 3 consecutive letters are distinct
    For example:
    is_happy(a) => False
    is_happy(aa) => False
    is_happy(abcd) => True
    is_happy(aabb) => False
    is_happy(adb) => True
    is_happy(xyy) => False
    """
\end{Verbatim}

%% file: Prompts/he_program_2.tex
\begin{Verbatim}[breaklines=true, breakanywhere=true]
def fix_spaces(text):
    """
    Given a string text, replace all spaces in it with underscores, 
    and if a string has more than 2 consecutive spaces, 
    then replace all consecutive spaces with - 
    
    fix_spaces("Example") == "Example"
    fix_spaces("Example 1") == "Example_1"
    fix_spaces(" Example 2") == "_Example_2"
    fix_spaces(" Example   3") == "_Example-3"
    """
\end{Verbatim}

%% file: Prompts/mbpp_program_prompt.tex
\begin{Verbatim}[breaklines=true, breakanywhere=true]
def sum_range_list(list1 : list, m : int, n : int) -> int:
    """
    Write a function to find the sum of numbers in a list within a range specified by two indices.
    """
\end{Verbatim}

%% file: Prompts/mbpp_program_2.tex
\begin{Verbatim}[breaklines=true, breakanywhere=true]
def diff_even_odd(list1 : list) -> int:
    """
    Write a function to find the difference of the first even and first odd number of a given list.
    """
\end{Verbatim}

%% file: Prompts/he_testcase_prompt.tex
\begin{Verbatim}[breaklines=true, breakanywhere=true]
def is_happy(s):
    """You are given a string s.
    Your task is to check if the string is happy or not.
    A string is happy if its length is at least 3 and every 3 consecutive letters are distinct
    For example:
    is_happy(a) => False
    is_happy(aa) => False
    is_happy(abcd) => True
    is_happy(aabb) => False
    is_happy(adb) => True
    is_happy(xyy) => False
    """

    pass # To-do: implement

# Check if is_happy works
assert is_happy(
\end{Verbatim}

%% file: Prompts/he_testcase_2.tex
\begin{Verbatim}[breaklines=true, breakanywhere=true]
def fix_spaces(text):
    """
    Given a string text, replace all spaces in it with underscores, 
    and if a string has more than 2 consecutive spaces, 
    then replace all consecutive spaces with - 
    
    fix_spaces("Example") == "Example"
    fix_spaces("Example 1") == "Example_1"
    fix_spaces(" Example 2") == "_Example_2"
    fix_spaces(" Example   3") == "_Example-3"
    """

    pass # To-do: implement

# Check if fix_spaces works
assert fix_spaces(
\end{Verbatim}

%% file: Prompts/mbpp_testcase_prompt.tex
\begin{Verbatim}[breaklines=true, breakanywhere=true]
def sum_range_list(list1 : list, m : int, n : int) -> int:
    """
    Write a function to find the sum of numbers in a list within a range specified by two indices.
    """
    pass # To-do: implement

# Check if sum_range_list works
assert sum_range_list(
\end{Verbatim}

%% file: Prompts/mbpp_testcase_2.tex
\begin{Verbatim}[breaklines=true, breakanywhere=true]
def diff_even_odd(list1 : list) -> int:
    """
    Write a function to find the difference of the first even and first odd number of a given list.
    """
    pass # To-do: implement

# Check if diff_even_odd works
assert diff_even_odd(
\end{Verbatim}

%% file: Prompts/HumanEval_spec_prompt_cushman.tex
\begin{Verbatim}[breaklines=true, breakanywhere=true]
# Problem 1

from typing import List

def filtered_even_integers(input_list: List[int]) -> List[int]:
    """ Given a list of integers, return a list that filters out the even integers.
    >>>  filtered_even_integers([1, 2, 3, 4])
    [1, 3]
    >>> filtered_even_integers([5, 4, 3, 2, 1])
    [5, 3, 1]
    >>> filtered_even_integers([10, 18, 20])
    []
    """
    # TODO
    pass

# Test 1

def test_filtered_even_integers(input_list: List()):
    """ Given an input `input_list`, test whether the function `filtered_even_integers` is implemented correctly.
    """
    # execute the function
    output_list = filtered_even_integers(input_list)

    # check if the output list only contains odd integers
    for integer in output_list:
        assert integer % 2 == 1
    # check if all the integers in the output list can be found in the input list
    for integer in output_list:
        assert integer in input_list

# run the testing function `test_filtered_even_integers` on 3 different input cases that satisfy the description
test_filtered_even_integers([31, 24, 18, 99, 1000, 523, 901])
test_filtered_even_integers([2, 4, 6, 8])
test_filtered_even_integers([500, 0, 302, 19, 7, 5])

# Problem 2

def repeat_vowel(input_str: str) -> str:
    """ Return a string where the vowels (`a`, `e`, `i`, `o`, `u`, and their capital letters) are repeated twice in place.
    >>> repeat_vowel('abcdefg')
    'aabcdeefg'
    >>> repeat_vowel('Amy Emily Uber')
    'AAmy EEmiily UUbeer'
    """
    # TODO
    pass

# Test 2

def test_repeat_vowel(input_str: str) :
    """ Given an input `input_str`, test whether the function `repeat_vowel` is implemented correctly.
    """
    # execute the function
    output_str = repeat_vowel(input_str)

    vowels = ['a', 'A', 'e', 'E', 'i', 'I', 'o', 'O', 'u', 'U']
    # check if the number of vowels in the  output string is doubled
    # First get the number of vowels in the input
    number_of_vowels_input = sum([input_str.count(vowel) for vowel in vowels])
    # Then get the number of vowels in the output
    number_of_vowels_output = sum([output_str.count(vowel) for vowel in vowels])
    assert number_of_vowels_input * 2 == number_of_vowels_output

# run the testing function `test_repeat_vowel` on 3 different input cases that satisfy the description
test_repeat_vowel('ABCDEabcdeABCDE YOUUOY')
test_repeat_vowel('I am a student')
test_repeat_vowel('sounds good to me')
\end{Verbatim}

%% file: Prompts/HumanEval_spec_prompt_davinci.tex
\begin{Verbatim}[breaklines=true, breakanywhere=true]
# Problem 1

from typing import List

def filtered_even_integers(input_list: List[int]) -> List[int]:
    """ Given a list of integers, return a list that filters out the even integers.
    >>>  filtered_even_integers([1, 2, 3, 4])
    [1, 3]
    >>> filtered_even_integers([5, 4, 3, 2, 1])
    [5, 3, 1]
    >>> filtered_even_integers([10, 18, 20])
    []
    """
    # TODO
    pass

# Test 1

def test_filtered_even_integers(input_list: List()):
    """ Given an input `input_list`, test whether the function `filtered_even_integers` is implemented correctly.
    """
    # execute the function
    output_list = filtered_even_integers(input_list)

    # check if the output list only contains odd integers
    for integer in output_list:
        assert integer % 2 == 1
    # check if all the integers in the output list can be found in the input list
    for integer in output_list:
        assert integer in input_list

# run the testing function `test_filtered_even_integers` on 3 different input cases that satisfy the description
test_filtered_even_integers([31, 24, 18, 99, 1000, 523, 901])
test_filtered_even_integers([2, 4, 6, 8])
test_filtered_even_integers([500, 0, 302, 19, 7, 5])

# Problem 2

def repeat_vowel(input_str: str) -> str:
    """ Return a string where the vowels (`a`, `e`, `i`, `o`, `u`, and their capital letters) are repeated twice in place.
    >>> repeat_vowel('abcdefg')
    'aabcdeefg'
    >>> repeat_vowel('Amy Emily Uber')
    'AAmy EEmiily UUbeer'
    """
    # TODO
    pass

# Test 2

def test_repeat_vowel(input_str: str) :
    """ Given an input `input_str`, test whether the function `repeat_vowel` is implemented correctly.
    """
    # execute the function
    output_str = repeat_vowel(input_str)

    vowels = ['a', 'A', 'e', 'E', 'i', 'I', 'o', 'O', 'u', 'U']
    # check if the number of vowels in the  output string is doubled
    # First get the number of vowels in the input
    number_of_vowels_input = sum([input_str.count(vowel) for vowel in vowels])
    # Then get the number of vowels in the output
    number_of_vowels_output = sum([output_str.count(vowel) for vowel in vowels])
    assert number_of_vowels_input * 2 == number_of_vowels_output

# run the testing function `test_repeat_vowel` on 3 different input cases that satisfy the description
test_repeat_vowel('ABCDEabcdeABCDE YOUUOY')
test_repeat_vowel('I am a student')
test_repeat_vowel('sounds good to me')

# Problem 3

def find_missing_number(nums: List[int]) -> int:
    """
    Given a list of n-1 integers in the range of 1 to n, find the one missing number.
    >>> find_missing_number([1, 2, 4, 6, 3, 7, 8])
    5
    >>> find_missing_number([5, 1, 4, 2])
    3
    """
    # TODO
    pass

# Test 3

def test_find_missing_number(nums: List[int]):
    """ Given an input `nums`, test whether the function `find_missing_number` is implemented correctly.
    """

    # execute the function
    output = find_missing_number(nums)

    n = len(nums) + 1
    # check if the output is an integer
    assert isinstance(output, int)
    # check if the output is in the range of 1 to n
    assert 1 <= output <= n
    # check if the output is the missing number
    assert output not in nums

# run the testing function `test_find_missing_number` on 3 different input cases that satisfy the description
test_find_missing_number([1, 3, 4, 6, 5, 7, 8])
test_find_missing_number([10, 9, 8, 7, 6, 5, 4, 3, 2])
test_find_missing_number([3, 2, 1, 6, 5, 4, 10, 9, 8])

# Problem 4

def find_kth_largest(nums: List[int], k: int) -> int:
    """
    Given an unsorted array of integers, find the kth largest element.
    >>> find_kth_largest([3, 2, 1, 5, 6, 4], 2)
    5
    """
    # TODO
    pass

# Test 4

def test_find_kth_largest(nums: List[int], k: int):
    """ Given an input `nums` and `k`, test whether the function `find_kth_largest` is implemented correctly.
    """
    # execute the function
    output = find_kth_largest(nums, k)

    # check if the output is an integer
    assert isinstance(output, int)
    # check if the output is in the input list
    assert output in nums
    # check if the output is the kth largest element
    assert output == sorted(nums)[-k]

# run the testing function `test_find_kth_largest` on 3 different input cases that satisfy the description
test_find_kth_largest([1, 10, 9, 2, 7, 6, -3, 4, 5, 8], 3)
test_find_kth_largest([100, 200, 900, 1000, 80, 101010], 5)
test_find_kth_largest([88, 131, 89, 125, 3, 7], 2)

# Problem 5

def reverse_substrings(s: str, indices: List[int]) -> str:
    """
    Given a string s and a list of integers representing starting and ending indices of substrings
    within s (inclusive), reverse each substring and return the modified string.
    >>> reverse_substrings('abcdefg', [1, 2, 4, 6])
    'acbdgfe'
    """
    # TODO
    pass

# Test 5

def test_reverse_substrings(s: str, indices: List[int]):
    """ Given an input `s` and `indices`, test whether the function `reverse_substrings` is implemented correctly.
    """
    # execute the function
    output = reverse_substrings(s, indices)

    # check if the function is implemented correctly
    # check if the output is a string
    assert isinstance(output, str)
    # check if the output is the same length as the input
    assert len(output) == len(s)
    # check if the output contains the same characters as the input
    assert set(output) == set(s)
    # check if all the substrings are reversed
    for i in range(0, len(indices), 2):
        assert output[indices[i]:indices[i+1]+1] == s[indices[i]:indices[i+1]+1][::-1]
    # check if all the other characters are the same
    for i in range(len(s)):
        # check if i in the indices
        if any(indices[i] <= i <= indices[i+1] for i in range(0, len(indices), 2)):
            continue
        assert output[i] == s[i]

# run the testing function `test_reverse_substrings` on 3 different input cases that satisfy the description
test_reverse_substrings('apple', [1, 2, 4, 5])
test_reverse_substrings('summerSpringWinterfall', [0, 5, 8, 9, 10, 15, 16, 20])
test_reverse_substrings('lkjhgfdqwert', [0, 3, 4, 7, 9, 11])
\end{Verbatim}

%% file: Prompts/MBPP_spec_prompt_cushman.tex
\begin{Verbatim}[breaklines=true, breakanywhere=true]
# Problem 1

from typing import List

def filtered_even_integers(input_list: List[int]) -> List[int]:
    """
    Given a list of integers, return a list that filters out the even integers.
    """
    # TODO
    pass

# Test 1

def test_filtered_even_integers(input_list: List()):
    """
    Given an input `input_list`, test whether the function `filtered_even_integers` is implemented correctly.
    """
    # execute the function
    output_list = filtered_even_integers(input_list)

    # check if the output list only contains odd integers
    for integer in output_list:
        assert integer % 2 == 1
    # check if all the integers in the output list can be found in the input list
    for integer in output_list:
        assert integer in input_list

# run the testing function `test_filtered_even_integers` on 3 different input cases that satisfy the description
test_filtered_even_integers([31, 24, 18, 99, 1000, 523, 901])
test_filtered_even_integers([2, 4, 6, 8])
test_filtered_even_integers([500, 0, 302, 19, 7, 5])

# Problem 2

def repeat_vowel(input_str: str) -> str:
    """
    Return a string where the vowels (`a`, `e`, `i`, `o`, `u`, and their capital letters) are repeated twice in place.
    """
    # TODO
    pass

# Test 2

def test_repeat_vowel(input_str: str) :
    """
    Given an input `input_str`, test whether the function `repeat_vowel` is implemented correctly.
    """
    # execute the function
    output_str = repeat_vowel(input_str)

    vowels = ['a', 'A', 'e', 'E', 'i', 'I', 'o', 'O', 'u', 'U']
    # check if the number of vowels in the  output string is doubled
    # First get the number of vowels in the input
    number_of_vowels_input = sum([input_str.count(vowel) for vowel in vowels])
    # Then get the number of vowels in the output
    number_of_vowels_output = sum([output_str.count(vowel) for vowel in vowels])
    assert number_of_vowels_input * 2 == number_of_vowels_output

# run the testing function `test_repeat_vowel` on 3 different input cases that satisfy the description
test_repeat_vowel('ABCDEabcdeABCDE YOUUOY')
test_repeat_vowel('I am a student')
test_repeat_vowel('sounds good to me')
\end{Verbatim}

%% file: Prompts/MBPP_spec_prompt_davinci.tex
\begin{Verbatim}[breaklines=true, breakanywhere=true]
# Problem 1

from typing import List

def filtered_even_integers(input_list: List[int]) -> List[int]:
    """
    Given a list of integers, return a list that filters out the even integers.
    """
    # TODO
    pass

# Test 1

def test_filtered_even_integers(input_list: List()):
    """
    Given an input `input_list`, test whether the function `filtered_even_integers` is implemented correctly.
    """
    # execute the function
    output_list = filtered_even_integers(input_list)

    # check if the output list only contains odd integers
    for integer in output_list:
        assert integer % 2 == 1
    # check if all the integers in the output list can be found in the input list
    for integer in output_list:
        assert integer in input_list

# run the testing function `test_filtered_even_integers` on 3 different input cases that satisfy the description
test_filtered_even_integers([31, 24, 18, 99, 1000, 523, 901])
test_filtered_even_integers([2, 4, 6, 8])
test_filtered_even_integers([500, 0, 302, 19, 7, 5])

# Problem 2

def repeat_vowel(input_str: str) -> str:
    """
    Return a string where the vowels (`a`, `e`, `i`, `o`, `u`, and their capital letters) are repeated twice in place.
    """
    # TODO
    pass

# Test 2

def test_repeat_vowel(input_str: str) :
    """
    Given an input `input_str`, test whether the function `repeat_vowel` is implemented correctly.
    """
    # execute the function
    output_str = repeat_vowel(input_str)

    vowels = ['a', 'A', 'e', 'E', 'i', 'I', 'o', 'O', 'u', 'U']
    # check if the number of vowels in the  output string is doubled
    # First get the number of vowels in the input
    number_of_vowels_input = sum([input_str.count(vowel) for vowel in vowels])
    # Then get the number of vowels in the output
    number_of_vowels_output = sum([output_str.count(vowel) for vowel in vowels])
    assert number_of_vowels_input * 2 == number_of_vowels_output

# run the testing function `test_repeat_vowel` on 3 different input cases that satisfy the description
test_repeat_vowel('ABCDEabcdeABCDE YOUUOY')
test_repeat_vowel('I am a student')
test_repeat_vowel('sounds good to me')

# Problem 3

def find_missing_number(nums: List[int]) -> int:
    """
    Given a list of n-1 integers in the range of 1 to n, find the one missing number.
    """
    # TODO
    pass

# Test 3

def test_find_missing_number(nums: List[int]):
    """
    Given an input `nums`, test whether the function `find_missing_number` is implemented correctly.
    """

    # execute the function
    output = find_missing_number(nums)

    n = len(nums) + 1
    # check if the output is an integer
    assert isinstance(output, int)
    # check if the output is in the range of 1 to n
    assert 1 <= output <= n
    # check if the output is the missing number
    assert output not in nums

# run the testing function `test_find_missing_number` on 3 different input cases that satisfy the description
test_find_missing_number([1, 3, 4, 6, 5, 7, 8])
test_find_missing_number([10, 9, 8, 7, 6, 5, 4, 3, 2])
test_find_missing_number([3, 2, 1, 6, 5, 4, 10, 9, 8])

# Problem 4

def find_kth_largest(nums: List[int], k: int) -> int:
    """
    Given an unsorted array of integers, find the kth largest element.
    """
    # TODO
    pass

# Test 4

def test_find_kth_largest(nums: List[int], k: int):
    """
    Given an input `nums` and `k`, test whether the function `find_kth_largest` is implemented correctly.
    """
    # execute the function
    output = find_kth_largest(nums, k)

    # check if the output is an integer
    assert isinstance(output, int)
    # check if the output is in the input list
    assert output in nums
    # check if the output is the kth largest element
    assert output == sorted(nums)[-k]

# run the testing function `test_find_kth_largest` on 3 different input cases that satisfy the description
test_find_kth_largest([1, 10, 9, 2, 7, 6, -3, 4, 5, 8], 3)
test_find_kth_largest([100, 200, 900, 1000, 80, 101010], 5)
test_find_kth_largest([88, 131, 89, 125, 3, 7], 2)

# Problem 5

def reverse_substrings(s: str, indices: List[int]) -> str:
    """
    Given a string s and a list of integers representing starting and ending indices of substrings
    within s (inclusive), reverse each substring and return the modified string.
    """
    # TODO
    pass

# Test 5

def test_reverse_substrings(s: str, indices: List[int]):
    """
    Given an input `s` and `indices`, test whether the function `reverse_substrings` is implemented correctly.
    """
    # execute the function
    output = reverse_substrings(s, indices)

    # check if the function is implemented correctly
    # check if the output is a string
    assert isinstance(output, str)
    # check if the output is the same length as the input
    assert len(output) == len(s)
    # check if the output contains the same characters as the input
    assert set(output) == set(s)
    # check if all the substrings are reversed
    for i in range(0, len(indices), 2):
        assert output[indices[i]:indices[i+1]+1] == s[indices[i]:indices[i+1]+1][::-1]
    # check if all the other characters are the same
    for i in range(len(s)):
        # check if i in the indices
        if any(indices[i] <= i <= indices[i+1] for i in range(0, len(indices), 2)):
            continue
        assert output[i] == s[i]

# run the testing function `test_reverse_substrings` on 3 different input cases that satisfy the description
test_reverse_substrings('apple', [1, 2, 4, 5])
test_reverse_substrings('summerSpringWinterfall', [0, 5, 8, 9, 10, 15, 16, 20])
test_reverse_substrings('lkjhgfdqwert', [0, 3, 4, 7, 9, 11])
\end{Verbatim}

%% file: Prompts/transformation.tex
Here we show how to transform the input problem to the prompt used for generating logical relations.
\paragraph{MBPP transformation}
First we parse the input problems from MBPP dataset and get the string representation of library imports, function name, function parameters, return type, and English problem description. We denote them as \texttt{imports}, \texttt{func\_name}, \texttt{parameter\_format}, \texttt{return\_type}, and \texttt{description} respectively and \texttt{problem\_number} is number of few-shot examples plus $1$.

Then we use the template shown in Figure~\ref{mbpp_input_output_template} and Figure~\ref{mbpp_logic_template} for input-output and logical relations respectively. The parsed string from the input problem would then be inserted to the placeholder accordingly.
\begin{figure}[H]
\begin{Verbatim}[breaklines=true, breakanywhere=true,frame=single]
# Problem 3

def {func_name}({", ".join(parameter_format)}) -> {return_type}:
    """
    {description}
    """
    pass # To-do: implement

# Test 3

def test_{func_name}(
\end{Verbatim}
\caption{Template for MBPP logical relation prompt}
\label{mbpp_logic_template}
\end{figure}

\begin{figure}[H]
\begin{Verbatim}[breaklines=true, breakanywhere=true,frame=single]
# Problem {problem_number}

{function_definition_with_description}
    # TODO
    pass

# Test {problem_number}
\end{Verbatim}
\caption{Template for MBPP input-output prompt}
\label{mbpp_input_output_template}
\end{figure}

Then, for the logical relations, we prepend the resulting string with the few shot example string shown in \ref{mbpp_fewshot_prompt}. For the input-output, we strip out the blank lines prefix if import is empty. 

\paragraph{HumanEval Transformation}
Similar to the above MBPP transformation,
we parse the input problems from HumanEval dataset and get the string representation of function definition plus English description and function name. We denote them as \texttt{function\_definition\_with\_description} and \texttt{func\_name}. Then we insert these into the template shown in Figure~\ref{humaneval_input_output_template} and Figure~\ref{humaneval_logic_template} for input-output and logical relations respectively.

\begin{figure}[H]
\begin{Verbatim}[breaklines=true, breakanywhere=true,frame=single]
{function_definition_with_description}
    pass

# Check if {func_name} works
assert {func_name}(
\end{Verbatim}
\caption{Template for HumanEval input-output prompt}
\label{humaneval_input_output_template}
\end{figure}

\begin{figure}[H]
\begin{Verbatim}[breaklines=true, breakanywhere=true,frame=single]
# Problem {problem_number}

{function_definition_with_description}
    # TODO
    pass

# Test {problem_number}
\end{Verbatim}
\caption{Template for HumanEval logical relation prompt}
\label{humaneval_logic_template}
\end{figure}
Finally, we prepend the resulting string with the few shot example string shown in \ref{humaneval_fewshot_prompt}.

%% file: aaai-anonymous-submission-latex-2024.bbl
\begin{thebibliography}{31}
\providecommand{\natexlab}[1]{#1}

\bibitem[{Acquaviva et~al.(2021)Acquaviva, Pu, Kryven, Wong, Ecanow, Nye,
  Sechopoulos, Tessler, and Tenenbaum}]{larc}
Acquaviva, S.; Pu, Y.; Kryven, M.; Wong, C.; Ecanow, G.~E.; Nye, M.~I.;
  Sechopoulos, T.; Tessler, M.~H.; and Tenenbaum, J.~B. 2021.
\newblock Communicating Natural Programs to Humans and Machines.
\newblock \emph{CoRR}, abs/2106.07824.

\bibitem[{Alur et~al.(2013)Alur, Bodik, Juniwal, Martin, Raghothaman, Seshia,
  Singh, Solar-Lezama, Torlak, and Udupa}]{alur2013syntax}
Alur, R.; Bodik, R.; Juniwal, G.; Martin, M.~M.; Raghothaman, M.; Seshia,
  S.~A.; Singh, R.; Solar-Lezama, A.; Torlak, E.; and Udupa, A. 2013.
\newblock \emph{Syntax-guided synthesis}.
\newblock IEEE.

\bibitem[{Austin et~al.(2021)Austin, Odena, Nye, Bosma, Michalewski, Dohan,
  Jiang, Cai, Terry, Le et~al.}]{austin2021program}
Austin, J.; Odena, A.; Nye, M.; Bosma, M.; Michalewski, H.; Dohan, D.; Jiang,
  E.; Cai, C.; Terry, M.; Le, Q.; et~al. 2021.
\newblock Program synthesis with large language models.
\newblock \emph{arXiv preprint arXiv:2108.07732}.

\bibitem[{Baldoni et~al.(2018)Baldoni, Coppa, D’elia, Demetrescu, and
  Finocchi}]{baldoni2018survey}
Baldoni, R.; Coppa, E.; D’elia, D.~C.; Demetrescu, C.; and Finocchi, I. 2018.
\newblock A survey of symbolic execution techniques.
\newblock \emph{ACM Computing Surveys (CSUR)}, 51(3): 1--39.

\bibitem[{Brown et~al.(2020)Brown, Mann, Ryder, Subbiah, Kaplan, Dhariwal,
  Neelakantan, Shyam, Sastry, Askell, Agarwal, Herbert{-}Voss, Krueger,
  Henighan, Child, Ramesh, Ziegler, Wu, Winter, Hesse, Chen, Sigler, Litwin,
  Gray, Chess, Clark, Berner, McCandlish, Radford, Sutskever, and Amodei}]{gpt}
Brown, T.~B.; Mann, B.; Ryder, N.; Subbiah, M.; Kaplan, J.; Dhariwal, P.;
  Neelakantan, A.; Shyam, P.; Sastry, G.; Askell, A.; Agarwal, S.;
  Herbert{-}Voss, A.; Krueger, G.; Henighan, T.; Child, R.; Ramesh, A.;
  Ziegler, D.~M.; Wu, J.; Winter, C.; Hesse, C.; Chen, M.; Sigler, E.; Litwin,
  M.; Gray, S.; Chess, B.; Clark, J.; Berner, C.; McCandlish, S.; Radford, A.;
  Sutskever, I.; and Amodei, D. 2020.
\newblock Language Models are Few-Shot Learners.
\newblock \emph{CoRR}, abs/2005.14165.

\bibitem[{Chen et~al.(2023)Chen, Zhang, Nguyen, Zan, Lin, Lou, and
  Chen}]{chen2022codet}
Chen, B.; Zhang, F.; Nguyen, A.; Zan, D.; Lin, Z.; Lou, J.-G.; and Chen, W.
  2023.
\newblock CodeT: Code Generation with Generated Tests.
\newblock In \emph{The Eleventh International Conference on Learning
  Representations}.

\bibitem[{Chen, Zaharia, and Zou(2023)}]{chen2023frugalgpt}
Chen, L.; Zaharia, M.; and Zou, J. 2023.
\newblock FrugalGPT: How to Use Large Language Models While Reducing Cost and
  Improving Performance.
\newblock arXiv:2305.05176.

\bibitem[{Chen et~al.(2021)Chen, Tworek, Jun, Yuan, Pinto, Kaplan, Edwards,
  Burda, Joseph, Brockman, Ray, Puri, Krueger, Petrov, Khlaaf, Sastry, Mishkin,
  Chan, Gray, Ryder, Pavlov, Power, Kaiser, Bavarian, Winter, Tillet, Such,
  Cummings, Plappert, Chantzis, Barnes, Herbert-Voss, Guss, Nichol, Paino,
  Tezak, Tang, Babuschkin, Balaji, Jain, Saunders, Hesse, Carr, Leike, Achiam,
  Misra, Morikawa, Radford, Knight, Brundage, Murati, Mayer, Welinder, McGrew,
  Amodei, McCandlish, Sutskever, and Zaremba}]{codex}
Chen, M.; Tworek, J.; Jun, H.; Yuan, Q.; Pinto, H. P. d.~O.; Kaplan, J.;
  Edwards, H.; Burda, Y.; Joseph, N.; Brockman, G.; Ray, A.; Puri, R.; Krueger,
  G.; Petrov, M.; Khlaaf, H.; Sastry, G.; Mishkin, P.; Chan, B.; Gray, S.;
  Ryder, N.; Pavlov, M.; Power, A.; Kaiser, L.; Bavarian, M.; Winter, C.;
  Tillet, P.; Such, F.~P.; Cummings, D.; Plappert, M.; Chantzis, F.; Barnes,
  E.; Herbert-Voss, A.; Guss, W.~H.; Nichol, A.; Paino, A.; Tezak, N.; Tang,
  J.; Babuschkin, I.; Balaji, S.; Jain, S.; Saunders, W.; Hesse, C.; Carr,
  A.~N.; Leike, J.; Achiam, J.; Misra, V.; Morikawa, E.; Radford, A.; Knight,
  M.; Brundage, M.; Murati, M.; Mayer, K.; Welinder, P.; McGrew, B.; Amodei,
  D.; McCandlish, S.; Sutskever, I.; and Zaremba, W. 2021.
\newblock Evaluating Large Language Models Trained on Code.

\bibitem[{Chen et~al.(2022)Chen, Ma, Wang, and Cohen}]{chen2022program}
Chen, W.; Ma, X.; Wang, X.; and Cohen, W.~W. 2022.
\newblock Program of Thoughts Prompting: Disentangling Computation from
  Reasoning for Numerical Reasoning Tasks.
\newblock arXiv:2211.12588.

\bibitem[{D'silva, Kroening, and Weissenbacher(2008)}]{d2008survey}
D'silva, V.; Kroening, D.; and Weissenbacher, G. 2008.
\newblock A survey of automated techniques for formal software verification.
\newblock \emph{IEEE Transactions on Computer-Aided Design of Integrated
  Circuits and Systems}, 27(7): 1165--1178.

\bibitem[{Fink and Bishop(1997)}]{fink1997property}
Fink, G.; and Bishop, M. 1997.
\newblock Property-based testing: a new approach to testing for assurance.
\newblock \emph{ACM SIGSOFT Software Engineering Notes}, 22(4): 74--80.

\bibitem[{Gao et~al.(2023)Gao, Madaan, Zhou, Alon, Liu, Yang, Callan, and
  Neubig}]{gao2023pal}
Gao, L.; Madaan, A.; Zhou, S.; Alon, U.; Liu, P.; Yang, Y.; Callan, J.; and
  Neubig, G. 2023.
\newblock PAL: Program-aided Language Models.
\newblock arXiv:2211.10435.

\bibitem[{Gulwani et~al.(2017)Gulwani, Polozov, Singh
  et~al.}]{gulwani2017program}
Gulwani, S.; Polozov, O.; Singh, R.; et~al. 2017.
\newblock Program synthesis.
\newblock \emph{Foundations and Trends{\textregistered} in Programming
  Languages}, 4(1-2): 1--119.

\bibitem[{Inala et~al.(2022)Inala, Wang, Yang, Codas, Encarnaci{\'o}n, Lahiri,
  Musuvathi, and Gao}]{inala2022fault}
Inala, J.~P.; Wang, C.; Yang, M.; Codas, A.; Encarnaci{\'o}n, M.; Lahiri,
  S.~K.; Musuvathi, M.; and Gao, J. 2022.
\newblock Fault-Aware Neural Code Rankers.
\newblock \emph{arXiv preprint arXiv:2206.03865}.

\bibitem[{Kadavath et~al.(2022)Kadavath, Conerly, Askell, Henighan, Drain,
  Perez, Schiefer, Dodds, DasSarma, Tran-Johnson et~al.}]{kadavath2022language}
Kadavath, S.; Conerly, T.; Askell, A.; Henighan, T.; Drain, D.; Perez, E.;
  Schiefer, N.; Dodds, Z.~H.; DasSarma, N.; Tran-Johnson, E.; et~al. 2022.
\newblock Language Models (Mostly) Know What They Know.
\newblock \emph{arXiv preprint arXiv:2207.05221}.

\bibitem[{Kuhn, Gal, and Farquhar(2023)}]{kuhn2023semantic}
Kuhn, L.; Gal, Y.; and Farquhar, S. 2023.
\newblock Semantic Uncertainty: Linguistic Invariances for Uncertainty
  Estimation in Natural Language Generation.
\newblock In \emph{The Eleventh International Conference on Learning
  Representations}.

\bibitem[{Lewkowycz et~al.(2022)Lewkowycz, Andreassen, Dohan, Dyer,
  Michalewski, Ramasesh, Slone, Anil, Schlag, Gutman-Solo
  et~al.}]{lewkowycz2022solving}
Lewkowycz, A.; Andreassen, A.; Dohan, D.; Dyer, E.; Michalewski, H.; Ramasesh,
  V.; Slone, A.; Anil, C.; Schlag, I.; Gutman-Solo, T.; et~al. 2022.
\newblock Solving quantitative reasoning problems with language models.
\newblock \emph{arXiv preprint arXiv:2206.14858}.

\bibitem[{Li et~al.(2022)Li, Choi, Chung, Kushman, Schrittwieser, Leblond,
  Eccles, Keeling, Gimeno, Lago et~al.}]{li2022competition}
Li, Y.; Choi, D.; Chung, J.; Kushman, N.; Schrittwieser, J.; Leblond, R.;
  Eccles, T.; Keeling, J.; Gimeno, F.; Lago, A.~D.; et~al. 2022.
\newblock Competition-level code generation with alphacode.
\newblock \emph{arXiv preprint arXiv:2203.07814}.

\bibitem[{Lin, Hilton, and Evans(2022)}]{lin2022teaching}
Lin, S.; Hilton, J.; and Evans, O. 2022.
\newblock Teaching Models to Express Their Uncertainty in Words.
\newblock \emph{arXiv preprint arXiv:2205.14334}.

\bibitem[{Manna and Waldinger(1979)}]{1702636}
Manna, Z.; and Waldinger, R. 1979.
\newblock Synthesis: Dreams → Programs.
\newblock \emph{IEEE Transactions on Software Engineering}, SE-5(4): 294--328.

\bibitem[{Ni et~al.(2023)Ni, Iyer, Radev, Stoyanov, Yih, Wang, and
  Lin}]{ni2023lever}
Ni, A.; Iyer, S.; Radev, D.; Stoyanov, V.; Yih, W.-t.; Wang, S.~I.; and Lin,
  X.~V. 2023.
\newblock LEVER: Learning to Verify Language-to-Code Generation with Execution.
\newblock \emph{arXiv preprint arXiv:2302.08468}.

\bibitem[{Nye et~al.(2021)Nye, Andreassen, Gur{-}Ari, Michalewski, Austin,
  Bieber, Dohan, Lewkowycz, Bosma, Luan, Sutton, and
  Odena}]{DBLP:journals/corr/abs-2112-00114}
Nye, M.~I.; Andreassen, A.~J.; Gur{-}Ari, G.; Michalewski, H.; Austin, J.;
  Bieber, D.; Dohan, D.; Lewkowycz, A.; Bosma, M.; Luan, D.; Sutton, C.; and
  Odena, A. 2021.
\newblock Show Your Work: Scratchpads for Intermediate Computation with
  Language Models.
\newblock \emph{CoRR}, abs/2112.00114.

\bibitem[{Peleg et~al.(2020)Peleg, Gabay, Itzhaky, and Yahav}]{10.1145/3428227}
Peleg, H.; Gabay, R.; Itzhaky, S.; and Yahav, E. 2020.
\newblock Programming with a Read-Eval-Synth Loop.
\newblock \emph{Proc. ACM Program. Lang.}, 4(OOPSLA).

\bibitem[{Platt et~al.(1999)}]{platt1999probabilistic}
Platt, J.; et~al. 1999.
\newblock Probabilistic outputs for support vector machines and comparisons to
  regularized likelihood methods.
\newblock \emph{Advances in large margin classifiers}, 10(3): 61--74.

\bibitem[{Pu et~al.(2020)Pu, Ellis, Kryven, Tenenbaum, and
  Solar-Lezama}]{pu2020program}
Pu, Y.; Ellis, K.; Kryven, M.; Tenenbaum, J.; and Solar-Lezama, A. 2020.
\newblock Program synthesis with pragmatic communication.
\newblock \emph{Advances in Neural Information Processing Systems}, 33:
  13249--13259.

\bibitem[{Shi et~al.(2022)Shi, Fried, Ghazvininejad, Zettlemoyer, and
  Wang}]{shi2022natural}
Shi, F.; Fried, D.; Ghazvininejad, M.; Zettlemoyer, L.; and Wang, S.~I. 2022.
\newblock Natural Language to Code Translation with Execution.
\newblock \emph{arXiv preprint arXiv:2204.11454}.

\bibitem[{Si et~al.(2022)Si, Gan, Yang, Wang, Wang, Boyd-Graber, and
  Wang}]{si2022prompting}
Si, C.; Gan, Z.; Yang, Z.; Wang, S.; Wang, J.; Boyd-Graber, J.; and Wang, L.
  2022.
\newblock Prompting gpt-3 to be reliable.
\newblock \emph{arXiv preprint arXiv:2210.09150}.

\bibitem[{Thoppilan et~al.(2022)Thoppilan, Freitas, Hall, Shazeer,
  Kulshreshtha, Cheng, Jin, Bos, Baker, Du, Li, Lee, Zheng, Ghafouri, Menegali,
  Huang, Krikun, Lepikhin, Qin, Chen, Xu, Chen, Roberts, Bosma, Zhou, Chang,
  Krivokon, Rusch, Pickett, Meier{-}Hellstern, Morris, Doshi, Santos, Duke,
  Soraker, Zevenbergen, Prabhakaran, Diaz, Hutchinson, Olson, Molina,
  Hoffman{-}John, Lee, Aroyo, Rajakumar, Butryna, Lamm, Kuzmina, Fenton, Cohen,
  Bernstein, Kurzweil, Aguera{-}Arcas, Cui, Croak, Chi, and Le}]{lambda}
Thoppilan, R.; Freitas, D.~D.; Hall, J.; Shazeer, N.; Kulshreshtha, A.; Cheng,
  H.; Jin, A.; Bos, T.; Baker, L.; Du, Y.; Li, Y.; Lee, H.; Zheng, H.~S.;
  Ghafouri, A.; Menegali, M.; Huang, Y.; Krikun, M.; Lepikhin, D.; Qin, J.;
  Chen, D.; Xu, Y.; Chen, Z.; Roberts, A.; Bosma, M.; Zhou, Y.; Chang, C.;
  Krivokon, I.; Rusch, W.; Pickett, M.; Meier{-}Hellstern, K.~S.; Morris,
  M.~R.; Doshi, T.; Santos, R.~D.; Duke, T.; Soraker, J.; Zevenbergen, B.;
  Prabhakaran, V.; Diaz, M.; Hutchinson, B.; Olson, K.; Molina, A.;
  Hoffman{-}John, E.; Lee, J.; Aroyo, L.; Rajakumar, R.; Butryna, A.; Lamm, M.;
  Kuzmina, V.; Fenton, J.; Cohen, A.; Bernstein, R.; Kurzweil, R.;
  Aguera{-}Arcas, B.; Cui, C.; Croak, M.; Chi, E.~H.; and Le, Q. 2022.
\newblock LaMDA: Language Models for Dialog Applications.
\newblock \emph{CoRR}, abs/2201.08239.

\bibitem[{Udupa et~al.(2013)Udupa, Raghavan, Deshmukh, Mador-Haim, Martin, and
  Alur}]{10.1145/2499370.2462174}
Udupa, A.; Raghavan, A.; Deshmukh, J.~V.; Mador-Haim, S.; Martin, M.~M.; and
  Alur, R. 2013.
\newblock TRANSIT: Specifying Protocols with Concolic Snippets.
\newblock \emph{SIGPLAN Not.}, 48(6): 287–296.

\bibitem[{Vaswani et~al.(2017)Vaswani, Shazeer, Parmar, Uszkoreit, Jones,
  Gomez, Kaiser, and Polosukhin}]{transformers}
Vaswani, A.; Shazeer, N.; Parmar, N.; Uszkoreit, J.; Jones, L.; Gomez, A.~N.;
  Kaiser, L.~u.; and Polosukhin, I. 2017.
\newblock Attention is All you Need.
\newblock In Guyon, I.; Luxburg, U.~V.; Bengio, S.; Wallach, H.; Fergus, R.;
  Vishwanathan, S.; and Garnett, R., eds., \emph{Advances in Neural Information
  Processing Systems}, volume~30. Curran Associates, Inc.

\bibitem[{Zhang et~al.(2022)Zhang, Yu, Hashimoto, Lewis, tau Yih, Fried, and
  Wang}]{zhang2022coder}
Zhang, T.; Yu, T.; Hashimoto, T.~B.; Lewis, M.; tau Yih, W.; Fried, D.; and
  Wang, S.~I. 2022.
\newblock Coder Reviewer Reranking for Code Generation.
\newblock arXiv:2211.16490.

\end{thebibliography}
